\documentclass[10pt,aps,prd,showpacs,superscriptaddress,twocolumn,nofootinbib]{revtex4}
\usepackage{verbatim} % allows to use \begin{comment} - \end{comment}
\usepackage{amsfonts}
\usepackage{amssymb}
\usepackage{amsmath}
\usepackage{MnSymbol}
\usepackage{graphicx}
\usepackage[caption=false]{subfig}
\usepackage{enumitem}
\usepackage{tikz}
\usetikzlibrary{decorations.pathmorphing,decorations.markings,calc,decorations.pathreplacing}
\usepackage{xcolor}

\date{\today}

\begin{document}

\title{Relativistic Quantum Communication:  Energy cost and channel capacities}
\author{Ian Bernardes Barcellos}\email{ian.barcellos@ufabc.edu.br}
\affiliation{Centro de Ci\^encias Naturais e Humanas,
Universidade Federal do ABC, 
Avenida dos Estados, 5001, 09210-580, Bangu,
{Santo Andr\'e, S\~ao Paulo, Brazil}}

\author{Andr\'e G.\ S.\ Landulfo}\email{andre.landulfo@ufabc.edu.br}
\affiliation{Centro de Ci\^encias Naturais e Humanas,
Universidade Federal do ABC, 
Avenida dos Estados, 5001, 09210-580, Bangu,
Santo Andr\'e, S\~ao Paulo, Brazil}

\begin{abstract}

We consider the communication of classical and quantum information between two arbitrary observers in asymptotically flat spacetimes (possibly containing black holes) and investigate what is the energy cost for such information transmission. By means of localized two-level quantum systems, sender and receiver can use a quantum scalar field as a communication channel. As we have already shown in a previous paper, such a channel has non-vanishing classical capacity as well as entanglement-assisted classical and quantum capacities. Here we will show that the change in the expectation value of the energy of the system during the communication process can be separated in: {\bf (i)} a contribution coming from the particle creation due to the change of the spacetime, {\bf (ii)} a contribution associated with the energy needed to switch-on/off each qubit, and {\bf (iii)} a term which comes from the communication process itself. For the quantum channel considered here, we show that the extra energy cost needed for communication vanishes. As a result, if one has already created a system of qubits for some specific task (e.g., quantum computation) one can also reliably convey information between its parts with no extra energy cost. We conclude the paper by illustrating the form of the channel capacities and energy contributions in two paradigmatic cases in Minkowski spacetime: {\bf (1)} sender and receiver in inertial motion and {\bf (2)} sender  in inertial motion while the receiver is uniformly accelerated.

\end{abstract}

\maketitle

%%%%%%%%%%%%%%%%%%%%%%%%%%%%%%%%%%%%%%%%%%%%%%%%%%%%%%%%%%%%%%%%%%%%%
\section{Introduction}
\label{sec:introduction}
%%%%%%%%%%%%%%%%%%%%%%%%%%%%%%%%%%%%%%%%%%%%%%%%%%%%%%%%%%%%%%%%%%%%

Quantify at which rate one can reliably convey information between two or more parts by means of a quantum communication channel is one of the major goals of quantum information theory~\cite{wilde, GT07}. Although the road to this end has a rich landscape already in the nonrelativistic context, it is only when relativity is taken into account that we can fully enjoy all its hues. The main reason is that relativity opens up the possibility of having non-trivial structures such as black hole event horizons, causal horizons caused by the relativistic relative motion of the parts conveying the information (or the expansion of spacetime itself), or even the presence of Cauchy horizons~\cite{wald84}. This has led several authors to analyze the communication process in relativistic settings with particular attention being paid to Minkowski~\cite{AM03,  BHTW10, LT13, MHM12, BHP12, CK10, JMK14, MM15, JRMK18, J17, HLL12, SAKM20,YASKM20}, Schwarzschild~\cite{HBK12, BA15, JACKM20}, or asymptotically flat cosmological spacetimes~\cite{MM, BGMM16,SM17}. 

In a recent paper~\cite{L16}, one of the authors analyzed a communication model using a bosonic quantum field as a communication channel which is suited to arbitrary observers communicating in any globally hyperbolic curved spacetime. In order to convey the information, both sender and receiver interact with the field, which can be in any quasi-free algebraic quantum state~\cite{wald94}, by means of localized two-level quantum systems (qubits). By analyzing such a quantum communication channel nonperturbatively, it was determined at which rate one can reliably transfer information between the two parts. It was shown that the channel has a nonvanishing classical capacity as well as entanglement-assisted classical and quantum capacities. 

However, it is not enough to know how much classical or quantum information can be transmitted between two parts. A related and equally relevant inquiry is: how much energy is needed to convey the information? This has important consequences not only for practical proposes (such as engineering communication networks) but because it may also shed light in one of the major open problems in semiclassical/quantum gravity nowadays, namely, what is the fate of the information which falls into a black hole. For instance, by studying the energy toll for information transmission one may be able to address if it is possible for all the information to come back at the end of the black hole evaporation process, if it needs to come back during its earlier stages, or even if the information gets destroyed/erased. (For a review of such an issue, see Ref.~\cite{UW} and references therein.) Although some investigation has been performed on the issue of how much energy is needed to convey information~\cite{B81}, they are usually restricted to Minkowski spacetime~\cite{JMK15, HSU15} or 1+1 spacetimes~\cite{J16, Wald19}. 

In order to gain a broader perspective on the subject, in the present paper we will use the communication channel developed in~\cite{L16} to analyze what is the energy cost to transmit information in general globally hyperbolic and asymptotically flat spacetimes (which may contain black holes). By means of the so-called null-surface quantization and its interplay with the usual quantization procedure, we will be able to analyze how the total energy of the system formed by the quantum field, $\phi$, and the two local qubits, $A$ and $B$, changes as it is evolved from the asymptotic past to the asymptotic future. It will be shown that such energy change can be separated in three terms: {\bf (i)} a contribution coming from the particle creation due to the change of the spacetime metric, {\bf (ii)} a contribution accounting for the energy needed to switch-on/off each qubit, and {\bf (iii)} a term which measures the extra energy cost  coming from  the communication process itself. For the communication channel being used, it will be proved that the contribution {\bf (iii)} coming from the convey of information vanishes. This shows that, once  one has already created the qubits, there is no extra energy cost in reliably transmitting information between the two parts.  

The paper is organized as follows. In Sec.~\ref{sec:Comm.Chann.} we will review the quantum communication model used here. In Sec.~\ref{sec:nullquant} we will develop the so-called null-surface quantization and relate it to the usual quantization procedure. In Section~\ref{sec:energy}, we will study how the energy of the total system--field $\phi$+qubits $AB$--changes when it is evolved from past to future null infinity and analyze what is the energy cost for communication.  Section~\ref{sec:Mink} will be used to illustrate the channel capacities and energy cost in two paradigmatic examples in Minkowski spacetime: two inertial observers and one inertial and the other uniformly accelerated. Section~\ref{sec:finalremarks} is reserved to our final remarks. We assume metric signature $(- + + +)$ and natural 
units in which $c=\hbar=G=1$ unless stated otherwise.

%%%%%%%%%%%%%%%%%%%%%%%%%%%%%%%%%%%%%%%%%%%%%%%%%%%%%%%%%%%%%%%%%%%%%
\section{The Communication Channel}
\label{sec:Comm.Chann.}
%%%%%%%%%%%%%%%%%%%%%%%%%%%%%%%%%%%%%%%%%%%%%%%%%%%%%%%%%%%%%%%%%%%%%

The quantum communication model we rely on describes the exchange of information between two arbitrary observers in any globally hyperbolic curved spacetime $\left(\mathcal{M},g\right)$ using a quantum scalar field $\phi$ as a communication channel. Here, $\mathcal{M}$ denotes the four-dimensional spacetime manifold and $g$ its Lorentzian metric. Let us consider a real free scalar field $\phi$ propagating in $\left(\mathcal{M},g\right)$. The spacetime can be foliated by Cauchy surfaces $\Sigma_t$ labeled by the real parameter $t$. The field is described by the action
\begin{equation}
  \label{eq:KG-action}
  S\equiv- \frac{1}{2}\int_\mathcal{M}\epsilon_\mathcal{M}\,
  ( \nabla_a\phi \nabla^a\phi + m^2 \phi^2 + \xi R \phi^2 ),
\end{equation}
where $\epsilon_\mathcal{M}=\sqrt{-\mathfrak{g}}dx^0\wedge \cdots \wedge dx^3$ is the spacetime volume 4-form, $m$ is the field mass, $\xi\in\mathbb{R}$, $R$ is the scalar curvature, $\nabla_a$ is the torsion-free covariant derivative compatible with the metric $g$, and $\mathfrak{g}\equiv\text{det}(g_{\mu\nu})$ in some arbitrary coordinate system. The extremization of the  action~\eqref{eq:KG-action} gives rise to the Klein-Gordon equation  
\begin{equation}
  \label{eq:KG-equation}
  (-\nabla^a\nabla_a + m^2+\xi R)\phi = 0.
\end{equation}

In the canonical quantization procedure, we promote the real field $\phi$ to an operator\footnote{Rigorously, an operator-valued distribution.} that satisfies the ``equal-time'' canonical commutation relations (CCR)
\begin{eqnarray}
\label{eq:CCR-unsmeared-1}
\, [\phi (t, {\bf x}), \phi (t, {\bf x}') ]_{\Sigma_t} 
& = & 
[\pi  (t, {\bf x}), \pi  (t, {\bf x}') ]_{\Sigma_t} = 0,
\\
\label{eq:CCR-unsmeared-2} 
\, [ \phi (t, {\bf x}), \pi  (t, {\bf x}')]_{\Sigma_t}
& = &  
i \delta^3 ({\bf x}, {\bf x}' ),
\end{eqnarray}
where $\mathbf{x}\equiv(x^1,x^2,x^3)$ are spatial coordinates on $\Sigma_t$ and $\pi(x)$ is the conjugate momentum defined as
\begin{equation}
  \label{eq:conjugate-momentum-definition}
  \pi\equiv\frac{\delta S}{\delta\dot{\phi}}\,,
\end{equation}
where $``\;\dot\;\;"\equiv\partial_t$. In addition, we may formally write the canonical Hamiltonian of the field as
\begin{equation}
  \label{eq:field-canonical-hamiltonian}
  H_\phi(t)\equiv\int_{\Sigma_t}d^3{\bf x} \,\left(\pi(t,\mathbf{x})\dot\phi(t,\mathbf{x})-\mathcal{L}[\phi,\nabla_a\phi]\right),
\end{equation}
with
\begin{equation}
     d^3{\bf x}\equiv dx^1\wedge d x^2 \wedge dx^3
\end{equation}
and
\begin{equation}
  \label{eq:field-lagrangian-density}
  \mathcal{L}[\phi,\nabla_a\phi]\equiv-\frac{1}{2}\sqrt{-\mathfrak{g}} \,( \nabla_a\phi \nabla^a\phi + {m}^2 \phi^2 + \xi R \phi^2)
\end{equation}
being the Lagrangian density.

To find a representation of the CCR, Eqs.~\eqref{eq:CCR-unsmeared-1} and~\eqref{eq:CCR-unsmeared-2}, we define an antisymmetric bilinear map $\sigma$ acting on the space $\mathcal{S}^\mathbb{C}$ of complex solutions of Eq.~\eqref{eq:KG-equation} as
\begin{equation}
  \label{eq:antisymmetric-bilinear-map}
  \sigma(\psi_1,\psi_2)\equiv\int_{\Sigma_t}\epsilon_\Sigma  \,n^a\left[\psi_2\nabla_a\psi_1-\psi_1\nabla_a\psi_2\right],
\end{equation}
where $\epsilon_\Sigma$ represents the proper-volume 3-form on the Cauchy surface $\Sigma_t$ and $n^a$ its future-directed normal unit vector. It allows us to define the Klein-Gordon product as
\begin{equation}
  \label{eq:KG-inner-product}
  \langle\psi_1,\psi_2\rangle\equiv -i\,\sigma(\overline{\psi}_1,\psi_2),
\end{equation}
and, although this product is not positive-definite on $\mathcal{S}^\mathbb{C}$, we may choose any subspace $\mathcal{H}\subset\mathcal{S}^\mathbb{C}$ (the so-called \textit{one-particle Hilbert space)} such that~\cite{wald94}: \textbf{(i)}~$\mathcal{S}^\mathbb{C}\simeq\mathcal{H}\bigoplus\overline{\mathcal{H}}$\footnote{For the sake of mathematical precision, we note that one must first suitably Cauchy-complete  $\mathcal{S}^\mathbb{C}$ for this decomposition to be valid.}; \textbf{(ii)}~the KG product is positive definite on $\mathcal{H}$, thus making $(\mathcal{H},\langle,\rangle)$ a Hilbert space\footnote{After its completion with respect to the norm induced by $\langle,\rangle$.}; \textbf{(iii)}~given any $u\in\mathcal{H}$ and $v\in\overline{\mathcal{H}}$, $\langle u,v\rangle=0$. Then, the Hilbert space that comprises the field states is defined as the symmetric Fock space $\mathfrak{F}_s(\mathcal{H})$ and the quantum field operator is formally defined as
\begin{equation}
  \label{eq:unsmeared-field-operator}
  \phi(t,\mathbf{x})\equiv\sum_j\left[u_j(t,\mathbf{x})a(\overline{u}_j)+\overline{u}_j(t,\mathbf{x})a^\dagger(u_j)\right],
\end{equation}
where $\{u_j\}$ form an orthonormal basis for $\mathcal{H}$ and $a(\overline{u})$ and $a^\dagger(v)$ are the usual annihilation and creation operators associated with the modes $u$ and $v$, respectively, which satisfy
\begin{equation}
  \label{eq:commutation-relation-annihilation-and-creation}
  \left[a(\overline{u}),a^\dagger(v)\right]=\langle u,v\rangle I,
\end{equation}
with $I$ being the identity operator on $\mathfrak{F}_s(\mathcal{H})$. The vacuum state associated with this representation of the CCR is the normalized vector $|0\rangle$ that satisfies $a(\overline{u})|0\rangle=0$ for every mode $u\in\mathcal{H}$.

In order to make it mathematically well-defined, the quantum field operator must be defined as an operator-valued distribution. To this end, let  $\mathcal{S}\subset\mathcal{S}^\mathbb{C}$ be the space of real solutions of Eq.~\eqref{eq:KG-equation} whose restriction to Cauchy surfaces have compact support; let $K:\mathcal{S}\rightarrow\mathcal{H}$ be the projection operator that takes the positive-norm part of any $\psi\in\mathcal{S}$; and define the map $E:C^\infty_0(\mathcal{M})\rightarrow\mathcal{S}$ acting on some \textit{test function}  $f\in C^\infty_0(\mathcal{M})$,  where $C^\infty_0(\mathcal{M})$ is the set of all smooth, compactly supported real functions on $\mathcal{M}$, as
\begin{equation}
  \label{eq:def-causal-propagator}
  Ef(x)\equiv Af(x)-Rf(x).
\end{equation}
Here, $Af$ and $Rf$ are the advanced and retarded solutions,  respectively, of the Klein-Gordon equation with source $f$. Hence, they satisfy
\begin{equation}
  \label{eq:KG-equation-source}
  P(Af) = P(Rf) = f,
\end{equation}
with $P\equiv(-\nabla^a\nabla_a + m^2+\xi R)$ representing the Klein-Gordon differential operator. Then, for each test function $f\in C^\infty_0(\mathcal{M})$, we define a \textit{smeared quantum field operator} by
\begin{equation}
  \label{eq:smeared-quantum-field-definition}
  \phi(f)\equiv i\left[a(\overline{KEf})-a^\dagger(KEf)\right],
\end{equation}
which satisfy the covariant version of the CCR,
\begin{equation}
  \label{eq:covariant-CCR}
  \left[\phi(f_1),\phi(f_2)\right]=-i\Delta(f_1,f_2)I,
\end{equation}
where
\begin{equation}
  \label{eq:def-nabla}
  \Delta(f_1,f_2)\equiv \int_\mathcal{M}\epsilon_\mathcal{M} f_1(x)Ef_2(x)
\end{equation}
and $f_1,f_2\in C^\infty_0(\mathcal{M})$. It is easy to see that Eq.~(\ref{eq:smeared-quantum-field-definition}) can be obtained by formally integrating Eq.~(\ref{eq:unsmeared-field-operator}) with the test function $f$, i.e., 
\begin{equation}
    \phi(f)=\int_\mathcal{M}\epsilon_\mathcal{M}\, \phi(x)f(x).
    \label{intfieldf}
\end{equation}

The above construction has the downside that there are infinitely many choices of $\mathcal{H}$ satisfying properties \textbf{(i)}-\textbf{(iii)} below Eq.~\eqref{eq:KG-inner-product} which are, in general, unitarily inequivalent. This issue can be avoided through the algebraic approach to quantum field theory (QFT)~\cite{wald94,KM15} in which the field quantization can be seen as a real linear map $\phi:f\in C^\infty_0(\mathcal{M})\rightarrow \phi(f)\in\mathcal{A}(\mathcal{M})$ between the space of test functions and an *-algebra $\mathcal{A}(\mathcal{M})$ (called \textit{algebra of observables}) such that
\begin{enumerate}
    \item $\phi(f)^*=\phi(f)$ for all  $f\in  C^\infty_0(\mathcal{M})$, i.e., the (smeared) field is Hermitian;
    \item $\phi(Pf)=0$, for all $f\in  C^\infty_0(\mathcal{M})$, i.e., the field satisfies the Klein-Gordon equation;
    \item $\left[\phi(f_1),\phi(f_2)\right]=-i\Delta(f_1,f_2)I,$ $f_1, f_2\in  C^\infty_0(\mathcal{M})$, i.e., the field satisfies the CCR;
    \item$\mathcal{A}(\mathcal{M})$ is algebraically generated by the identity $I$ and the $\phi(f)$'s, $f\in  C^\infty_0(\mathcal{M})$.
\end{enumerate}
In the algebraic approach, a quantum state is defined as a complex linear functional $\omega:\mathcal{A}(\mathcal{M})\rightarrow\mathbb{C}$ which satisfies $\omega(A^*A)\geq0$ for all $A\in\mathcal{A}(\mathcal{M})$ and $\omega(I)=1$. The so-called Gelfand-Naimark-Segal~(GNS) construction~\cite{wald94,KM15} ensures that every algebraic state $\omega$ can be realized as a vector on a Hilbert space together with a representation of the algebra of observables.

In this work, we will focus on a particular class of states: the \textit{quasifree states}, defined as follows. Given a real inner product $\mu:\mathcal{S}\times\mathcal{S}\rightarrow\mathbb{R}$ satisfying
\begin{equation}
   \label{eq:quasifree-state-condition}
   |\sigma(\varphi_1,\varphi_2)|^2\leq4\mu(\varphi_1,\varphi_1) \mu(\varphi_2,\varphi_2),
\end{equation}
for all $\varphi_1,\varphi_2\in\mathcal{S}$, we define a quasifree state $\omega_\mu$ associated with $\mu$ by the relation
\begin{equation}
    \label{eq:quasifree-state-definition}
    \omega_\mu\left[e^{i\phi(f)}\right]\equiv e^{-\mu(Ef,Ef)/2},
\end{equation}
for all $f\in C^\infty_0(\mathcal{M})$. We can extend $\omega_\mu$ to act on all observables of $\mathcal{A}(\mathcal{M})$ by using  
\begin{equation}
    \omega_\mu\left[\phi(f_1)\cdots \phi(f_n)\right]\!\equiv\! \left.(-i)^n\partial^n_{t_1\cdots t_n}\left[e^{it_1\phi(f_1)}\cdots e^{it_n\phi(f_n)}\right]\right|_{{\bf t} ={\bf 0}},
\end{equation}
where ${\bf t}\equiv (t_1,\cdots, t_n)$, together with linearity and continuity.

Now that the field quantization procedure has been introduced, we present the communication scheme studied here. Suppose that two observers, Alice and Bob, want to use the quantum field $\phi$ to communicate with each other. We consider Alice's and Bob's trajectories to be arbitrary and we consider the field to be initially in some quasifree state $\omega_\mu$\footnote{Actually, the results from this section apply to any algebraic state $\omega$  which satisfies $\omega\left[e^{i\phi(f)}\right]\in \mathbb{R}^+$. This condition includes the vacuum states, n-particle states, as well as KMS (thermal) states.}. Each observer possesses a two-level gapless quantum system that may interact with the quantum field. The two-dimensional Hilbert spaces associated with Alice's and Bob's qubits are denoted by $\mathcal{H}_A$ and $\mathcal{H}_B$, respectively.

The communication setup is as follows: Alice wants to transmit classical or quantum information to Bob, and for that, she prepares her qubit in some initial quantum state $\rho_{-\infty}^A$ and switches-on its interaction with the field for a finite time $\Delta t_A$ (measured in the parameter $t$). To measure the information imprinted by Alice on the field's state, Bob initially prepares his qubit in a suitable state~$\rho_{-\infty}^B$. He then switches-on its interaction with the field for a finite time interval $\Delta t_B$, but only after Alice has switched-off her qubit interaction. Such communication setup is implemented by means of the Hamiltonian 
\begin{equation}
  \label{eq:total-hamiltonian}
  H(t)\equiv H_\phi(t) + H_{\mathrm{int}}(t),
\end{equation}
where $H_\phi$ is the field Hamiltonian in Eq.~\eqref{eq:field-canonical-hamiltonian} and $H_\mathrm{int}$ is the Hamiltonian that describes the interaction between the qubits and the field which, in the interaction picture, is given by
\begin{equation}
  \label{eq:interaction-hamiltonian}
  H_\mathrm{int}^\mathrm{I} (t)\equiv \sum_{j}\epsilon_j(t)\int_{\Sigma_t}d^3{\bf x} \sqrt{-\mathfrak{g}} \; \psi_j(t,{\bf x}) \phi(t,{\bf x}) \otimes\sigma^{\rm z}_j,
\end{equation}
where $j=A,B$ with $A$ and $B$ labelling  Alice and Bob qubit, respectively. Here, $\sigma_j^\mathrm{z}$ is one of the Pauli matrices $\left\{\sigma^\mathrm{x}_j,\sigma^\mathrm{y}_j,\sigma_j^\mathrm{z}\right\}$ associated with qubit $j$, $\psi_j(t,\mathbf{x})$ is a smooth real function satisfying $\psi|_{\Sigma_t} \in C_0^\infty\left(\Sigma_t\right)$ for all $t$ which models the range of interaction between the qubit $j$ and the field (i.e., the interaction occurs only at some vicinity of each qubit worldline), and $\epsilon_j(t)$ is a smooth and compactly-supported real {\em coupling function} modeling the finite-time coupling of the qubit $j$ with the field. Each coupling function has support
\begin{equation}
  \label{eq:support-coupling-function}
  \mathrm{supp}\;\epsilon_j=\left[T_j^\mathrm{i},T_j^\mathrm{f}\right],
\end{equation}
where $T_j^\mathrm{i}$ and $T_j^\mathrm{f}$ represent the time (with respect to the parameter $t$) in which each qubit interaction with the field is switched-on and -off, respectively. Here, we denote $\Delta t_j\equiv {T_j^\mathrm{f}-T_j^\mathrm{i} }$ and assume $T_B^\mathrm{i}\geq T_A^\mathrm{f}$ (i.e., Bob measurement will be performed after Alice imprinted her information on the field state).

The interaction between each qubit and the field, given by Eq.~\eqref{eq:interaction-hamiltonian}, is very similar to the Unruh-DeWitt model. However, we assumed that the two levels of each qubit have the same (zero) energy. This assumption allows us to calculate the evolution operator of the system  and trace out the field degrees of freedom in a nonperturbative manner, thus making this model interesting to investigate both the causality in the information exchange process as well as the communication between parts lying in early and future asymptotic spacetime regions. We note that one could also have given an energy gap $2\,\delta_j$ for each qubit $j$ by adding  $H_j=\delta_j\sigma_j^\mathrm{z}$ to the total Hamiltonian in Eq.~\eqref{eq:total-hamiltonian}. This would change it  to  
\begin{equation}
  \label{eq:total-hamiltonian-with-gaps}
  H=H_\phi+H_A+H_B+H_\mathrm{int},
\end{equation}
but would keep the interaction Hamiltonian in the interaction picture, Eq.~\eqref{eq:interaction-hamiltonian}, unchanged. Hence, all the results we have just described would remain the same.

The interaction-picture time-evolution operator, associated with the foliation $\Sigma_t$, can be written as the time-ordered expression
\begin{equation}
  \label{eq:evolution-operator-definition}
  U\equiv T\exp\left[-i\int_{-\infty}^{\infty}dt\,H^\mathrm{I}_\mathrm{int}(t)\right].
\end{equation}
As shown in \cite{L16}, Eq.~\eqref{eq:evolution-operator-definition} can be computed exactly and it is given by
\begin{equation}
  \label{eq:evolution-operator-closed-form}
  U=e^{i\Xi}e^{-i\phi(f_A)\otimes \sigma^{\mathrm{z}}_A}e^{-i\phi(f_B)\otimes \sigma^{\mathrm{z}}_B}e^{-i\Delta(f_A,f_B)\sigma^{\mathrm{z}}_A\otimes\sigma^{\mathrm{z}}_B},
\end{equation}
where $\Xi$ is the c-number
\begin{equation*}
  \Xi\equiv\frac{1}{2}\sum_j\int_{-\infty}^\infty dt \; \epsilon_j(t)\int_{-\infty}^{t} \; dt' \epsilon_j(t') \Delta_{j}(t,t'),
\end{equation*}
with
\begin{equation*}
  \Delta_{j}(t,t')\!\equiv \!\!\!\int_{\Sigma_t}\!\!\!\!\!\!d^3{\mathbf{x}}\sqrt{-\mathfrak{g}} \!\!\int_{\Sigma_{t'}}\!\!\!\!\!\!\!d^3{\mathbf{x'}} \sqrt{-\mathfrak{g}'}\psi_j(t,{\mathbf{x}})\Delta(x,x')\psi_j(t',{\mathbf{x'}}),
\end{equation*}
and we recall that $[\phi(x),\phi(x')]\equiv-i\Delta(x,x')I$ is the unsmeared version of Eq.~\eqref{eq:covariant-CCR}. Additionally, we have defined
\begin{equation}
  \label{eq:qubit-funtion}
  f_j(t,\mathbf{x})\equiv \epsilon_j(t)\psi_j(t,\mathbf{x}),
\end{equation}
which is a compactly supported function on $\mathcal{M}$ carrying all information about the interaction of qubit $j$ with the field.

The initial state of the two-qubit+field system is ${\rho_{-\infty}\equiv\rho^A_{-\infty}\otimes\rho^B_{-\infty}\otimes\rho_\omega}$, where $\rho_\omega$ is the density operator associated with the initial quasifree state, $\omega_\mu$, of the field. Using the unitary evolution operator in Eq.~\eqref{eq:evolution-operator-closed-form}, we can evolve $\rho_{-\infty}$ to obtain the system's state after the communication process, $\rho_{+\infty}=U\rho_{-\infty}U^\dagger$. Additionally, we can trace-out the field and Alice's qubit degrees of freedom, obtaining the state of Bob's qubit after the communication process has finished:
\begin{equation}
  \label{eq:qubit-B-final-state-definition}
  \rho^B\equiv\mathrm{tr}_{A,\phi}\left(U\,\rho^A_{-\infty}\otimes\rho^B_{-\infty}\otimes\rho_\omega\,U^\dagger\right).
\end{equation}
As shown in \cite{L16}, Eq.~\eqref{eq:qubit-B-final-state-definition} can be written in the form
\begin{equation}
  \label{eq:quantum-communication-map}
  \rho^B=\mathcal{E}\left(\rho^A_{-\infty}\right),
\end{equation}
where $\mathcal{E}$ is a linear, completely-positive, and trace-preserving (CPTP) quantum map that relates the initial state of Alice's qubit (which has the information that will be conveyed) to the final state of Bob's qubit (which will be measured by him in order to retrieve the message). In other words, $\mathcal{E}$ is the quantum map that describes the communication channel used between Alice and Bob. It depends on their trajectories, on the spacetime geometry, and  on both the initial state of the field and of Bob's qubit. The explicit form of Eq.~\eqref{eq:quantum-communication-map} as well as the details of the calculations can be found in~\cite{L16}. It is worth pointing out, however, that the initial state $\rho_{-\infty}^B$ should not be arbitrarily chosen: since $\sigma^\mathrm{z}_B$ commutes with the total Hamiltonian \eqref{eq:total-hamiltonian-with-gaps}, $\sigma^\mathrm{z}_B$ is conserved and thus its eigenvalues cannot be used to recover any information transmitted by Alice. Nevertheless, it can be shown that some states will maximize the signalling amplitudes between Alice and Bob, e.g., $\rho^B_{-\infty}\equiv|y_+\rangle_B{}_B\langle y_+|$, where $|y_+\rangle_B$ satisfies $\sigma_B^\mathrm{y}|y_+\rangle_B=|y_+\rangle_B$. With this choice of $\rho^B_{-\infty}$, it can be shown that this quantum channel has a classical capacity (i.e., the maximum rate at which classical bits can be reliably transmitted) given by
\begin{equation}
    \label{eq:classical-channel-capacity}
    \begin{aligned}
        C(\mathcal{E}) = &\, H\left(\frac{1}{2}+\frac{\nu_B}{2}|\cos[2\Delta(f_A,f_B)]|\right)\\ 
         - & \,H\left(\frac{1}{2}+\frac{\nu_B}{2}\right),
    \end{aligned}
\end{equation}
where $H(x)\equiv-x\log_2{x}-(1-x)\log_2{(1-x)}$ is the Shannon entropy and
\begin{equation}
  \label{eq:nu-B-def}
  \nu_B\equiv\omega_\mu\left[e^{i\phi(2f_B)}\right]=e^{-2\mu(KEf_B,KEf_B)}.
\end{equation}

On the other hand, since this channel is entanglement-breaking, its quantum channel capacity (i.e., the rate at which qubits can be reliably transmitted) is
\begin{equation}
  \label{eq:quantum-channel-capacity}
  Q(\mathcal{E})=0.
\end{equation}

One could also define protocols for sending both classical and quantum
information when Alice and Bob initially
have access to an unlimited supply of entanglement. In this case, one can define classical (or quantum) entanglement-assisted channel capacities which measure the maximum rate at which classical information (or qubits) can be reliably sent through the channel. As shown in~\cite{L16}, the classical $C_{ea}(\mathcal{E})$ and quantum $Q_{ea}(\mathcal{E})$ entanglement-assisted capacities are related to the classical channel capacity~\eqref{eq:classical-channel-capacity} by
\begin{equation}
    \label{eq:entanglement-assisted-channel-capacities}
    C_{ea}(\mathcal{E})=2Q_{ea}(\mathcal{E})=C(\mathcal{E}).
\end{equation}
Thus, it is not worth using entanglement in order to try to increase
the classical capacity of this channel. On the other hand, when prior entanglement is shared between Alice and Bob, it is possible to convey qubits through this channel at maximum rate $Q_{ea}(\mathcal{E})$, in contrast with the unassisted case in Eq.~\eqref{eq:quantum-channel-capacity}.

%%%%%%%%%%%%%%%%%%%%%%%%%%%%%%%%%%%%%%%%%%%%%%%%%%%%%%%%%%%%%%%%%%%%%%
\section{Quantization on Null Surfaces}
\label{sec:nullquant}
%%%%%%%%%%%%%%%%%%%%%%%%%%%%%%%%%%%%%%%%%%%%%%%%%%%%%%%%%%%%%%%%%%%%%%

The so-called quantization on null-surfaces provides the formulation of a QFT restricted to 3-dimensional null submanifolds such as black hole horizons~\cite{MP03}, asymptotic infinities~\cite{M06,DMP06}, and cosmological horizons~\cite{DMP09}. In this section, we will build a quantum field theory restricted to a special class of null hypersurfaces. Then, under a few assumptions, we show how one can relate the ordinary QFT presented in Sec.~\ref{sec:Comm.Chann.} to the algebra of operators defined in the asymptotic past and future null infinities as well as causal horizons.

Let $\mathfrak{h}$ be a 3-dimensional null hypersurface satisfying:
\begin{enumerate}[label=\textbf{(\arabic*)}]
  \item $\mathfrak{h}$ is diffeomorphic to $\mathbb{R}\times\Gamma$, where $\Gamma$ is a two-dimensional spacelike submanifold of $\mathcal{M}$;
  \item there exists coordinates $(\Omega,\lambda,s^1,s^2)$ on $\mathcal{M}$ such that:
  \begin{enumerate}
      \item $s\equiv (s^1,s^2)$ are coordinates of $\Gamma$;
      \item $\mathfrak{h}=\{p\in\mathcal{M}\,|\,\Omega(p)=0\}$ and $d\Omega\neq0$ at $\mathfrak{h}$;
      \item the restriction of the metric to $\mathfrak{h}$ takes the form
    \begin{equation}
      g\big|_\mathfrak{h} = -\gamma^2\,\left(d\Omega \otimes d\lambda + d\lambda \otimes  d\Omega\right)+h_\Gamma
      \label{eq:metric-restricted-to-horizon}
    \end{equation}
    where $h_\Gamma$ is the metric induced by $g$ on $\Gamma$ and $\gamma\in\mathbb{R}$.
  \end{enumerate}
\end{enumerate}
It follows from conditions {\bf (1)} and {\bf (2)} above that $(\lambda,s)$ defines a coordinate system for $\mathfrak{h}$ and that the curves $\lambda\rightarrow(\lambda,s)$, defined for fixed $s$, are the null generators of $\mathfrak{h}$. From now on, we will refer to $\mathfrak{h}$ generically as ``the horizon'' (although it can also be describing past or future infinity). 

To parallel the usual QFT construction we have presented in Sec.~\ref{sec:Comm.Chann.}, let us define the ``solution space'' on $\mathfrak{h}$ as
\begin{equation}
  S^\mathbb{C}_\mathfrak{h}\equiv\left\{\mathrm{smooth}\;\psi:\mathfrak{h}\rightarrow\mathbb{C}\,\big|\,\psi,\partial_\lambda\psi\in L^2(\mathfrak{h},d\lambda\wedge\epsilon_\Gamma)\right\},
\end{equation}
where $\epsilon_\Gamma$ is the natural volume element on $\Gamma$ and $L^2(\mathfrak{h},d\lambda\wedge\epsilon_\Gamma)$ is the space of square-integrable functions on $\mathfrak{h}$ with respect to the measure $d\lambda\wedge\epsilon_\Gamma$. Similarly to Eq.~\eqref{eq:antisymmetric-bilinear-map}, we define the sympletic product on $\mathcal{S}^\mathbb{C}_\mathfrak{h}$ as
\begin{equation}
  \sigma_\mathfrak{h}(\psi_1,\psi_2) \equiv \int_\mathfrak{h} d\lambda\wedge \epsilon_\Gamma\left[\psi_2\partial_\lambda \psi_1-\psi_1\partial_\lambda\psi_2\right],
  \label{eq:symplectic-form-null-surface}
\end{equation}
which again allows us to define the Klein-Gordon inner-product on $\mathcal{S}^\mathbb{C}_\mathfrak{h}$ by
\begin{equation}
  \langle\psi_1,\psi_2\rangle_\mathfrak{h}\equiv-i\,\sigma_\mathfrak{h}(\overline{\psi_1},\psi_2).
  \label{eq:KG-product-null-surface}
\end{equation}

Now, let us note that Eq.~\eqref{eq:metric-restricted-to-horizon} is invariant under translations $\lambda\rightarrow\lambda+a$, with $a\in\mathbb{R}$. We can explore this translation symmetry on $\lambda$ to choose a preferable representation of the CCR. For this purpose, let us first define the \textit{positive-frequency projection operator} $K$ acting on $\psi\in\mathcal{S}^\mathbb{C}_\mathfrak{h}$ by
\begin{equation}
  K\psi(\lambda,s)\equiv\frac{1}{\sqrt{2\pi}}\int_{\mathbb{R}^+}dE\,e^{-i E \lambda}\,\widetilde{\psi}(E,s),
  \label{eq:projection-operator-null-surface}
\end{equation}
where
\begin{equation}
  \widetilde{\psi}(E,s)\equiv\frac{1}{\sqrt{2\pi}}\int_\mathbb{R}d\lambda\,e^{i E \lambda}\,\psi(\lambda,s)
\end{equation}
is the Fourier transform of $\psi$ with respect to $\lambda$. Then, the one-particle Hilbert space is defined as
\begin{equation}
  \mathcal{H}_\mathfrak{h}\equiv\overline{\left\{K\psi\,|\,\psi\in\mathcal{S}^\mathbb{C}_\mathfrak{h}\right\}}  \label{eq:one-particle-space-null-surface},
\end{equation}
where the closure is with respect to the norm induced by the product in Eq.~\eqref{eq:KG-product-null-surface} (which is positive definite on~$\mathcal{H}_\mathfrak{h}$). It is easy to see that the horizon one-particle Hilbert space $\mathcal{H}_\mathfrak{h}$ in Eq.~\eqref{eq:one-particle-space-null-surface} satisfies properties \textbf{(i)}-\textbf{(iii)} below Eq.~\eqref{eq:KG-inner-product}.

Having established the one-particle Hilbert space $H_\mathfrak{h}$, we can now follow the same procedure described in  Sec.~\ref{sec:Comm.Chann.} to build the bosonic Fock space $\mathfrak{F}_s(\mathcal{H_\mathfrak{h}})$ and the creation/annihilation operators associated with each mode $u\in\mathcal{H}_\mathfrak{h}$. Then, the \textit{horizon smeared quantum field} can be defined as
\begin{equation}
  \phi^{(\mathfrak{h})}(\psi)\equiv i \left[a(\overline{K\psi})-a^\dagger(K\psi)\right],
  \label{eq:smeared-quantum-field-null-surface}
\end{equation}
for every  $\psi\in\mathcal{S}_\mathfrak{h}\subset \mathcal{S}_\mathfrak{h}^{\mathbb{C}}$ with 
\begin{equation}
    \mathcal{S}_\mathfrak{h} \equiv\{\psi\in\mathcal{S}^\mathbb{C}_\mathfrak{h}\,|\,\psi\mathrm{\;is\;real}\}.
\end{equation}
The horizon algebra of observables, $\mathcal{A}(\mathfrak{h})$, is generated by the identity operator ${I:\mathfrak{F}_s(\mathcal{H}_\mathfrak{h})\rightarrow\mathfrak{F}_s(\mathcal{H}_\mathfrak{h})}$ and the set of field operators ${\{\phi(\psi)\,|\,\psi\in \mathcal{S}_\mathfrak{h}\}}$.

It will be useful to write a basis for the one-particle space $\mathcal{H}_\mathfrak{h}$. To this end, let $\{\varphi_\alpha\}_{\alpha\in \Lambda}\subset L^2(\Gamma,\epsilon_\Gamma)$ be an orthonormal basis for $L^2(\Gamma,\epsilon_\Gamma)$ with respect to some measure $d\mu(\alpha)$ on the set $\Lambda$ of quantum numbers $\alpha$. Hence, every ${\psi\in L^2(\Gamma,\epsilon_\Gamma)}$ can be written as
\begin{equation}
  \psi(s)=\int_{\Lambda} d\mu(\alpha)\,\tilde{\psi}(\alpha) \,\varphi_\alpha(s)
\end{equation}
for some function $\tilde{\psi}(\alpha)$, with $\varphi_\alpha,\varphi_\beta$ satisfying
\begin{equation}
  \int_\Gamma \epsilon_\Gamma\,\overline{\varphi_\alpha(s)}\,\varphi_\beta(s)=\delta_\mu(\alpha,\beta),
\end{equation}
where $\delta_\mu$ is the Dirac distribution relative to the measure $d\mu(\alpha)$. Then, define the set of modes $\{u_{E\alpha}\}\subset \mathcal{H}_\mathfrak{h}$ as 
\begin{equation}
  u_{E\alpha}(\lambda,s)\equiv \frac{1}{\sqrt{4\pi E}} e^{-i E\lambda}\varphi_\alpha(s),\;\;\;E>0,
  \label{eq:basis-modes-null-surface}
\end{equation}
which satisfy 
\begin{equation}
     \langle u_{E\alpha},u_{E', \alpha'}\rangle_\mathfrak{h}=\delta(E-E')\delta_\mu(\alpha, \alpha')
\end{equation}
and thus, form a orthonormal basis for the one-particle Hilbert space $\mathcal{H}_\mathfrak{h}$. They allow us to define annihilation operators, $a_{E\alpha}\equiv a\left(\overline{u_{E\alpha}}\right)$,  and write the \textit{horizon unsmeared quantum field operator} as
\begin{equation}
    \phi^{(\mathfrak{h})}(\lambda,s)\equiv \int_{\mathbb{R^+}}dE\int_\Lambda d\mu(\alpha) \left[\,u_{E\alpha}a_{E\alpha}+\mathrm{H.c.} \,\right],
  \label{eq:unsmeared-quantum-field-null-surface}
\end{equation}
which satisfy the commutation relation
\begin{equation}
  \left[\phi^{(\mathfrak{h})}(\lambda,s),\partial_\lambda\phi^{(\mathfrak{h})}(\lambda',s')\right]=\frac{i}{2}\delta(\lambda-\lambda')\delta_\Gamma(s-s').
  \label{eq:unsmeared-commutation-relation-horizon}
\end{equation}

It's worth noting that the smeared and unsmeared quantum fields, Eqs.~\eqref{eq:smeared-quantum-field-null-surface} and~\eqref{eq:unsmeared-quantum-field-null-surface}, are related by
\begin{equation}
  \begin{aligned}
    \phi^{(\mathfrak{h})}(\psi)&= \sigma_\mathfrak{h}(\psi,\phi^\mathfrak{h}) \\
              & =2\int_\mathfrak{h} d\lambda\wedge \epsilon_\Gamma\,\partial_\lambda \psi(\lambda,s) \phi^\mathfrak{h}(\lambda,s) \\
              & = \int_\mathfrak{h} 2 d\psi\wedge\epsilon_\Gamma\,\phi^\mathfrak{h}(\lambda,s),
  \end{aligned}
  \label{eq:smearing-horizon}
\end{equation}
from which we see that the correct way to smear this field is with forms. This is because there is no natural volume element on $\mathfrak{h}$ (the induced metric is degenerate).

Let us now discuss the application of the null-surface quantization and its relation to the ordinary QFT presented in Sec.~\ref{sec:Comm.Chann.}. Suppose that the spacetime $(\mathcal{M},g)$ is asymptotically flat with future null infinity $\mathcal{I}^+$, possibly containing a future causal horizon $\mathfrak{h}^+$ (e.g., the event horizon of a black hole). The future null infinity $\mathcal{I}^+$ is a 3-dimensional null hypersurface which satisfies the properties \textbf{(1)}-\textbf{(2)} defined at the beginning of this section with $\Gamma=\mathbb{S}^2$ and $\lambda \equiv u$ being the so-called ``retarded time''. Similarly, the causal horizon $\mathfrak{h}^+$ is a 3-dimensional null hypersurface which satisfies the same properties but with $\lambda\equiv v$ being the so-called ``advanced time''. Thus, we can apply the quantization procedure introduced in this section to both surfaces $\mathcal{I}^+$ and $\mathfrak{h}^+$ (if present) and build the field algebras $\mathcal{A}(\mathcal{I}^+)$ and $\mathcal{A}(\mathfrak{h}^+)$, respectively.

Let us take now $\mathfrak{h}\equiv\mathfrak{h}^+\cup\mathcal{I}^+$ as the union of future null infinity and the future causal horizon  and let $\widetilde{\mathcal{M}}\equiv I^-(\mathfrak{h})$ be the asymptotically flat region outside the horizon, where $I^-(A)$ indicates the chronological past of a subset $A\subset \mathcal{M}$. For the sake of simplicity, from now on, we will restrict our analysis to minimally-coupled massless fields.  Suppose we have constructed a quantum field theory in $\widetilde{\mathcal{M}}$ following the steps of Sec.~\ref{sec:Comm.Chann.}, obtaining an algebra of observables $\mathcal{A}(\widetilde{\mathcal{M}})$ (called the \textit{bulk algebra}). Since we are dealing with a massless field, all the information carried by the field will be ``imprinted'' on $\mathfrak{h}$. Thus, we expect that: {\bf (BB1)} every solution $\psi\in \mathcal{S}$ of the Klein-Gordon equation in $\widetilde{\mathcal{M}}$ which has compact support on Cauchy surfaces can be extended by continuity to (unique) functions $\psi^{\mathcal{I}^+}\in \mathcal{S}^{\mathbb{C}}_{\mathcal{I}^+}$ on $\mathcal{I}^+$ and  $\psi^{\mathfrak{h}^+}\in \mathcal{S}^{\mathbb{C}}_{\mathfrak{h}^+}$ on $\mathfrak{h}^+$. Moreover, $\psi, \psi^{\mathcal{I}^+},$ and $\psi^{\mathfrak{h}^+}$ should satisfy 
\begin{equation}
 \mathbf{(BB2)} \;\; \sigma(\psi_1,\psi_2)=\sigma_{\mathcal{I}^+}(\psi_1^{\mathcal{I}^+},\psi_2^{\mathcal{I}^+})+\sigma_{\mathfrak{h}^+}(\psi_1^{\mathfrak{h}^+},\psi_2^{\mathfrak{h}^+}),
\end{equation}
where the LHS is defined in Eq.~\eqref{eq:antisymmetric-bilinear-map} and the horizon bilinear products in the RHS are defined in Eq.~\eqref{eq:symplectic-form-null-surface}. By using {\bf (BB1)} and {\bf (BB2)} above, each operator in $A(\widetilde{\mathcal{M}})$ can then be mapped with an operator in ${\mathcal{A}(\mathfrak{h})\equiv\mathcal{A}(\mathcal{I}^+)\otimes\mathcal{A}(\mathfrak{h}^+)}$ by the identification
\begin{equation}
  \phi(f)\rightarrow\phi^{(\mathfrak{h})}(Ef^\mathfrak{h}),\;\;\;\;\forall f\in C^\infty_0\left(\widetilde{\mathcal{M}}\right).
\end{equation}
An algebraic state ${\omega:\mathcal{A}(\widetilde{M})\rightarrow\mathbb{R}^+}$ induces the state ${\omega_\mathfrak{h}:\mathcal{A}(\mathcal{\mathfrak{h})}\rightarrow\mathbb{R}^+}$ on $\mathfrak{h}$ through the identification
\begin{equation}
  \omega_\mathfrak{h}\left[\phi^{(\mathfrak{h})}(Ef^\mathfrak{h})\right]\equiv \omega\left[\phi(f)\right]\;,\;\;\;\;\forall\phi^{(\mathfrak{h})}(Ef^\mathfrak{h})\in\mathcal{A}(\mathfrak{h}).
\end{equation}
By following a completely analogous procedure, one can also relate the bulk algebra to the algebra defined at past null infinity, $\mathcal{I}^-$.

It is important to note that, although we will restrict ourselves to massless and minimally-coupled real scalar fields, the above relation between between $\mathcal{A}(\widetilde{M})$ and $\mathcal{A}(\mathfrak{h})$ will always exists provided that the field in question satisfies conditions {\bf (BB1)} and {\bf (BB2)}~\cite{DMP09}.

%%%%%%%%%%%%%%%%%%%%%%%%%%%%%%%%%%%%%%%%%%%%%%%%%%%%%%%%%%%%%%%%%%%%
\section{Energy Cost for the Transmission of Information}
\label{sec:energy}
%%%%%%%%%%%%%%%%%%%%%%%%%%%%%%%%%%%%%%%%%%%%%%%%%%%%%%%%%%%%%%%%%%%%

In Sec.~\ref{sec:Comm.Chann.}, we have discussed a communication channel that allows the transmission of information between two arbitrary observers in a globally hyperbolic spacetime $(\mathcal{M}, g)$. Now, we turn our attention to investigate the energy cost involved in this communication process when $(\mathcal{M}, g)$ is asymptotically flat with past and future null infinities given by $\mathcal{I}^-$ and $\mathcal{I}^+,$ respectively. Our goal will be to analyze the total energy variation of the two-qubit+field system between early and late times. We recall that the initial state of the  system is given by
\begin{equation}
  \rho_{-\infty}\equiv\rho^A_{-\infty}\otimes\rho^B_{-\infty}\otimes\rho_\omega,
\end{equation}
where $\rho_\omega$ is the density operator associated with some initial quasi-free field state $\omega_\mu$ and $\rho^j_{-\infty}$ is the initial state of qubit $j=A,B$. When the communication process finishes, the final state of the two-qubits+field system is
\begin{equation}
  \rho_{+\infty}\equiv U\rho_{-\infty}U^\dagger,
  \label{eq:initial-state}
\end{equation}
where $U$ is the evolution operator given by Eq.~\eqref{eq:evolution-operator-closed-form}. As a result, the total energy variation of the system is formally written as
\begin{equation}
  \Delta E \equiv \langle H(+\infty) \rangle_{\rho_{+\infty}} - \langle H(-\infty)\rangle_{\rho_{-\infty}},
  \label{eq:definition-energy-difference}
\end{equation}
with $H(t)$ defined in Eq.~\eqref{eq:total-hamiltonian}. As the interaction time of each qubit with the field is finite, the interaction Hamiltonian vanishes for $t\rightarrow\pm\infty$ and thus, Eq.~\eqref{eq:definition-energy-difference} can be cast as
\begin{equation}
  \begin{aligned}
    \Delta E = \mathrm{tr}\left(H_\phi(+\infty) U\rho_{-\infty}U^\dagger \right)-\mathrm{tr}\left(H_\phi(-\infty)\rho_{-\infty}\right).
  \end{aligned}
  \label{eqDeltaE}
\end{equation}

As in Sec.~\ref{sec:Comm.Chann.}, define  $\mathfrak{h}\equiv\mathfrak{h}^+\cup\mathcal{I}^+$ or $\mathfrak{h}\equiv\mathcal{I}^+$, depending on whether there is a future causal horizon $\mathfrak{h}^+$ or not. Now, let us restrict our attention to the globally-hyperbolic region $\widetilde{\mathcal{M}}=I^-\left(\mathfrak{h}\right)$ outside the horizon and let us foliate it with Cauchy surfaces $\Sigma_t$ such that $\Sigma_{t\rightarrow -\infty}=\mathcal{I}^-$ and  $\Sigma_{t\rightarrow \infty}=\mathfrak{h}$. By using the identification between the algebra  $\mathcal{A}\left(\widetilde{\mathcal{M}}\right)$ with the algebras $\mathcal{A}\left(\mathfrak{h}\right)$ and  $\mathcal{A}\left(\mathcal{I}^-\right)$,  we can cast Eq.~\eqref{eqDeltaE} as 
\begin{eqnarray}
    \Delta E & =& \mathrm{tr}\left(H^{(\mathfrak{h})}_\phi\,U^{(\mathfrak{h})}\rho_{-\infty}^\mathfrak{h}U^{(\mathfrak{h})\,\dagger}\right)-\mathrm{tr}\left(H_\phi^{(\mathcal{I}^-)}\rho_{-\infty}^{(\mathcal{I}^-)}\right) \nonumber \\
             & =& \mathrm{tr}\left(U^{(\mathfrak{h})\,\dagger}H^{(\mathfrak{h})}_\phi\,U^{(\mathfrak{h})}\rho_{-\infty}^\mathfrak{h}\right)-\mathrm{tr}\left(H_\phi^{(\mathcal{I}^-)}\rho_{-\infty}^{(\mathcal{I}^-)}\right),\nonumber\\
  \label{eq:delta-E-horizon}
\end{eqnarray}
with $H^{(\mathfrak{h})}_\phi$ $\left(H_\phi^{(\mathcal{I}^-)}\right)$ and  $\rho_{-\infty}^{\left(\mathfrak{h}\right)}$ $\left(\rho_{-\infty}^{(\mathcal{I}^-)}\right)$ being the horizon field Hamiltonian and the state induced by $\rho_{-\infty}$ at $\mathfrak{h}$ $\left(\mathcal{I}^-\right)$, respectively. Similarly, $U^{(\mathfrak{h})}$ is the evolution operator~\eqref{eq:evolution-operator-closed-form} written using the algebra $\mathcal{A}\left(\mathfrak{h}\right)$, i.e., we have used the identification $\phi(f_j) \rightarrow \phi^{(\mathfrak{h})}(Ef_j^\mathfrak{h})$.
\begin{comment}
\begin{equation}
  U^{(\mathfrak{h})} =e^{i\Xi}e^{-i\phi^{(\mathfrak{h})}(Ef_A^\mathfrak{h})\otimes \sigma^{\mathrm{z}}_A}e^{-i\phi^{(\mathfrak{h})}(Ef_B^\mathfrak{h})\otimes \sigma^{\mathrm{z}}_B}e^{-i\Delta(f_A,f_B)\sigma^{\mathrm{z}}_A\otimes\sigma^{\mathrm{z}}_B}.
\end{equation}
\end{comment}

The field Hamiltonian at $X=\mathcal{I}^-, \mathfrak{h}$ can be written as
\begin{equation}
  \begin{aligned}
    H^{X}_\phi & = \int_{X} d\lambda_X \wedge\epsilon_{\Gamma_X}\; T^{X}_{ab}k^ak^b \\
    & =\int_{X} d\lambda_X\wedge\epsilon_{\Gamma_X} \left[\partial_{\lambda_X} \phi^{X}\right]^2.
  \end{aligned}
  \label{eq:energy-horizon-expression}
\end{equation}
where $\Gamma_X$ is the spacelike 2-surface transverse to the null  generators of X,
\begin{equation}
    T^{X}_{ab}\equiv \nabla_{(a}\phi^{X}\nabla_{b)}\phi^{X}-\frac{1}{2}g_{ab}\nabla_c\phi^{X}\nabla^c\phi^{X}
    \label{TX}
\end{equation}
is the stress-energy-momentum tensor at $X$ of the massless KG field,  and $k^a\equiv (\partial_{\lambda_X})^a$ is the vector field tangent to the affinely-parametrized null generators of $X$ whose affine parameter is given by $\lambda_{\mathcal{I}^-}= v$ or $\lambda_{\mathfrak{h}}=\lambda$ whenever $X=\mathcal{I}^-$ or $X=\mathfrak{h}$, respectively.

Let us evaluate Eq.~\eqref{eq:delta-E-horizon} by steps. For this purpose, we first define
\begin{equation}
    U^{(\mathfrak{h})}_j\equiv e^{-i\phi^{(\mathfrak{h})}(Ef_j^\mathfrak{h})\otimes\sigma_j^\mathrm{z}},\;\;j=A,B,
    \label{Uj}
\end{equation}
and use Eq.~\eqref{eq:evolution-operator-closed-form}, together with the identification between field algebras in $\mathfrak{h}$ and $\widetilde{\mathcal{M}}$, to write 
\begin{equation}
  {U^{(\mathfrak{h})}}^\dagger H_\phi^{(\mathfrak{h})}U^{(\mathfrak{h})}= \\
  {U^{(\mathfrak{h})}_B}^\dagger {U^{(\mathfrak{h})}_A}^\dagger\,H_\phi^{(\mathfrak{h})}U^{(\mathfrak{h})}_A U^{(\mathfrak{h})}_B.
  \label{eqHhorizonU}
\end{equation}
Next, by using Eqs.~(\ref{eq:unsmeared-commutation-relation-horizon}) and~(\ref{Uj}) together with the relation
\begin{equation}
  e^{\mathfrak{a}}\mathfrak{b}e^{-\mathfrak{a}}=\mathfrak{b}+[\mathfrak{a},\mathfrak{b}],
\end{equation}
valid when $\left[[\mathfrak{a},\mathfrak{b}],\mathfrak{a}\right]=\left[[\mathfrak{a},\mathfrak{b}],\mathfrak{b}\right]= 0$, we can write
\begin{equation}
    {U^{(\mathfrak{h})}_j}^\dagger \,\partial_\lambda \phi^{\mathfrak{h}}\,U_j^{(\mathfrak{h})}=\partial_\lambda \phi^{\mathfrak{h}}- \partial_\lambda Ef_j^\mathfrak{h}\,\sigma^\mathrm{z}_j,
    \label{UHU}
\end{equation}
where we recall that $j=A, B$. Now, using Eqs.~(\ref{Uj}) and~(\ref{UHU}) in Eq.~(\ref{eqHhorizonU}), we obtain 
\begin{equation}
    {U^{(\mathfrak{h})}}^\dagger \,\partial_\lambda \phi^{\mathfrak{h}}\,U^{(\mathfrak{h})}=\partial_\lambda \phi^{\mathfrak{h}}-  \sum_{j=A,B}\partial_\lambda Ef_j^\mathfrak{h}\,\sigma^\mathrm{z}_j.
    \label{UHU2}
\end{equation}
By using Eqs.~(\ref{eq:energy-horizon-expression}) and~(\ref{UHU2}) we can cast the evolved Hamiltonian on $\mathfrak{h}$ as 
\begin{eqnarray}
&& U^{(\mathfrak{h})\dagger} H_\phi^{(\mathfrak{h})}U^{(\mathfrak{h})}
 =\; H_\phi^{(\mathfrak{h})}\nonumber \\  
 &-& 2 \sum_{j=A,B}\int_\mathfrak{h}d\lambda \wedge \epsilon_\Gamma \partial_\lambda Ef_j^\mathfrak{h} \partial_\lambda \phi^\mathfrak{h}\otimes\sigma_j^\mathrm{z} \nonumber  \\
 &+& \sum_{i,j=A,B}\int_\mathfrak{h}d\lambda \wedge \epsilon_\Gamma \partial_\lambda Ef_i^\mathfrak{h} \partial_\lambda Ef_j^\mathfrak{h}\sigma^\mathrm{z}_i\otimes \sigma^\mathrm{z}_j.
  \label{eq:evolution-hamiltonian-horizon-final}
\end{eqnarray}
%%%%%%%%
Finally, by substituting  Eq.~\eqref{eq:evolution-hamiltonian-horizon-final} in Eq.~\eqref{eq:delta-E-horizon} we can write the energy variation as
\begin{equation}
  \Delta E=W_\phi+W_A+W_B+W_{AB},
\end{equation}
where
\begin{equation}
  W_\phi\equiv\mathrm{tr}\left(H^{(\mathfrak{h})}_\phi\,\rho^{(\mathfrak{h})}_{\omega}\right)-\mathrm{tr}\left(H^{(\mathcal{I}^-)}_\phi\,\rho^{(\mathcal{I}^-)}_{\omega}\right),
  \label{eq:W-phi}
\end{equation}
\begin{equation}
  W_j\equiv \int_\mathfrak{h}d\lambda\wedge\epsilon_\Gamma \left(\partial_\lambda Ef^\mathfrak{h}_j\right)^2,
  \label{eq:W-j}
\end{equation}
\begin{equation}
  W_{AB}\equiv 2\left[\int_\mathfrak{h}d\lambda\wedge\epsilon_\Gamma\,\left(\partial_\lambda Ef^\mathfrak{h}_A\right)\left(\partial_\lambda Ef^\mathfrak{h}_B\right)\right]\langle\sigma_A^\mathrm{z}\rangle_{\rho^A_{-\infty}}\langle\sigma_B^\mathrm{z}\rangle_{\rho^B_{-\infty}},
  \label{eq:W-AB}
\end{equation}
and we have used that, for any quasi-free state $\omega$, $$\langle\partial_\lambda \phi^\mathfrak{h} \rangle_\omega\equiv {\rm tr}\left(\rho^{\mathfrak{h}}_\omega \partial_\lambda \phi^\mathfrak{h}\right)=0.$$

Note that we have separated the energy variation into three parts. The first one, $W_\phi$, is the contribution to the energy that arises from the particle creation due to the change in the spacetime metric. It depends only on the field state and spacetime metric and has nothing to do with the presence of Alice and Bob. A difficulty we face now is that some sort of renormalization of the field energy operator $H_\phi$ is needed. For this purpose, we will restrict ourselves to the so-called Hadamard states, for which a general renormalization procedure is possible~\cite{wald94}. By noting that any state that is Hadamard in some open neighborhood of a Cauchy surface is Hadamard everywhere~\cite{FSW78}, we can see that the spacetime evolution preserves the renormalizability of the state. As a result, we can see that, for Hadamard states, Eq.~(\ref{eq:W-phi}) (and, thus, Eq.~(\ref{eqDeltaE}) as $W_\phi$ is the only contribution to $\Delta E$ where divergences appear) is well-defined and gives finite results.  

The second contribution, $W_A+W_B$, depends independently on each qubit interaction with the field. This contribution is due to the work necessary to switch-on/off each qubit, and it depends on their trajectories, coupling constants, as well as on spacetime parameters. 

The third contribution, $W_{AB}$, measures the extra energy cost arising from the communication process itself. It depends on the initial state of each qubit, on the spacetime metric, and on the relative motion between Alice and Bob. We note that, by integrating by parts and using Eq.~(\ref{eq:symplectic-form-null-surface}), we can write
\begin{equation}
 2 \int_\mathfrak{h}d\lambda\wedge\epsilon_\Gamma\,\left(\partial_\lambda Ef^\mathfrak{h}_A\right)\left(\partial_\lambda Ef^\mathfrak{h}_B\right)=\sigma_{\mathfrak{h}}\left(Ef^\mathfrak{h}_A, \partial_\lambda Ef^\mathfrak{h}_B\right),
\end{equation}
which, by using Eqs.~(\ref{eq:unsmeared-commutation-relation-horizon}) and~(\ref{eq:smearing-horizon}), enable us to cast Eq.~(\ref{eq:W-AB}) as
\begin{equation}
    W_{AB}=\left\langle i\left[\phi^{(\mathfrak{h})}(Ef^\mathfrak{h}_A),\phi^{(\mathfrak{h})}(\partial_\lambda Ef_B)\right] \right\rangle_{\omega}\langle\sigma_A^\mathrm{z}\rangle_{\rho^A_{-\infty}}\langle\sigma_B^\mathrm{z}\rangle_{\rho^B_{-\infty}}
\end{equation}
As expected, we can see that $W_{AB}$ vanishes if Alice's and Bob's qubits interact with the field in causally disconnected regions of the spacetime. More interestingly, we can make the $W_{AB}$ contribution to identically vanish  with a convenient choice of $\rho^B_{-\infty}$. Recall that Alice encodes the information she wants to convey in her qubit's initial state $\rho^{A}_{-\infty}$. On the other hand, we are free to choose the initial state of Bob's qubit. The choice $\rho^B_{-\infty}\equiv|y+\rangle_B{}_B\langle y_+|$, for example, leads to $W_{AB}=0$ while it maximizes the channel capacities. This shows that one can convey arbitrary amounts of information through this quantum channel without extra energy costs.

%%%%%%%%%%%%%%%%%%%%%%%%%%%%%%%%%%%%%%%%%%%%%%%%%%%%%%%%%%%%%%%%%%%%%
\section{Two Paradigmatic Examples}
\label{sec:Mink}
%%%%%%%%%%%%%%%%%%%%%%%%%%%%%%%%%%%%%%%%%%%%%%%%%%%%%%%%%%%%%%%%%%%%%

We now illustrate the results presented in the previous sections with two paradigmatic examples in Minkowski spacetime. Let us begin with the field quantization, following the steps presented in Sec~\ref{sec:Comm.Chann.}. Consider a free and massless scalar field $\phi$ propagating in the Minkowski spacetime $(\mathbb{R}^4,\eta)$. Let $(t,x,y,z)\in\mathbb{R}^4$ denote global inertial Cartesian coordinates and let us denote the spatial coordinates as $\mathbf{x}\equiv(x,y,z)$. The Klein-Gordon equation is simply
\begin{equation}
  \Box \phi = 0,
  \label{eq:minkowski-KG-equation}
\end{equation}
with $\Box\equiv -\sum_{\mu,\nu}\eta^{\mu \nu}\partial_\mu\partial_\nu$,  $\eta\equiv \sum_{\mu,\nu}\eta_{\mu \nu}dx^\mu\otimes dx^\nu$, and 
\begin{equation}
\eta=-dt\otimes dt + dx\otimes dx +   dy\otimes dy + dz\otimes dz.    \label{eta}
\end{equation} 
Let $\mathcal{S}^\mathbb{C}$ be the space of complex solutions of Eq.~\eqref{eq:minkowski-KG-equation} with compact-support initial data and consider the antisymmetric bilinear map~\eqref{eq:antisymmetric-bilinear-map}, which takes the form
\begin{equation}
  \sigma(\psi_1,\psi_2) = \int_{\Sigma_{t=0}} d^3\mathbf{x}\left[\psi_2\partial_t\psi_1-\psi_1\partial_t\psi_2\right],
  \label{eq:KG-product-minkowski}
\end{equation}
where 
\begin{equation}
    \Sigma_{t={\rm cte}}\equiv\left\{{(t,{\bf x})\in \mathbb{R}^4| t={\rm cte}}\right\}.
    \label{SigmaInert}
\end{equation} 
We choose as the one-particle Hilbert space $\mathcal{H}$ the space spanned by the positive frequency parts,  with respect to the inertial time $t$, of solutions in $\mathcal{S}^\mathbb{C}$ Cauchy-completed with the norm induced by the Klein-Gordon inner product~\eqref{eq:KG-inner-product}. One then builds the bosonic Fock space $\mathfrak{F}_s(\mathcal{H})$ as usual, to represent the space of field states and define the field operators via Eq.~\eqref{eq:smeared-quantum-field-definition}. This is the standard CCR representation in Minkowski spacetime associated with inertial observers and we will refer to its vacuum state, $|0_M\rangle$, as the inertial (or Minkowski) vacuum state.

Using the Green functions of the D'alambertian operator $\Box$, one can show that the map $E:C^\infty_0(\mathcal{M})\rightarrow \mathcal{S}$ defined in Eq.~\eqref{eq:def-causal-propagator} takes the form~\cite{F89}
\begin{equation}
  Ef(x') =\int \epsilon_{\mathcal{M}} \,f(x)\,E(x',x)
  \label{eq:causal-operator-minkowski}
\end{equation}
with 
\begin{multline}
  E(x,x')\equiv\frac{1}{4\pi |\mathbf{x}-\mathbf{x'}|}\left[\phantom{\big|}\delta\left(t-t'-|\mathbf{x}-\mathbf{x'}|\right)\right. \\
  \left. - \,\delta\left(t-t'+ |\mathbf{x}-\mathbf{x'}|\right)\phantom{\big|}\right].
\end{multline}

For later use, it will be useful to consider the standard \textit{positive-frequency modes}
\begin{equation}
  u_\mathbf{k}(t,\mathbf{x})\equiv\frac{1}{4\pi^{\frac{3}{2}}|\mathbf{k}|^\frac{1}{2}}\, e^{-i|\mathbf{k}|t}e^{i\mathbf{k\cdot x}}\;,\;\;\;\mathbf{k}\in\mathbb{R}^3,
  \label{eq:inertial-modes}
\end{equation}
which comprises a complete basis for the one-particle Hilbert space $\mathcal{H}$.

Now that we have chosen a representation for the CCR in Minkowski spacetime, let us analyze the effects of the field state as well as the state of motion of both Alice and Bob in the communication process.

\subsection{Inertial sender and receiver}

We consider first the following scenario: suppose Alice is at rest at the origin of our inertial coordinate system and wants to transmit some information to Bob, which is at rest at the spatial position $\mathbf{x}=(L,0,0)$ (thus at rest relative to Alice and separated by a spatial distance $L$). For simplicity, we consider that both are equipped with pointlike detectors. To avoid divergences, we consider that the interactions of each qubit with the field are switched-on/off continuously. We have seen that Eq.~\eqref{eq:qubit-funtion} carries all the information about the qubit interaction with the field. Applying it to  Alice's qubit+field interaction gives 
\begin{equation}
  f_A(t,\mathbf{x})=\epsilon_Ac_A(t)\delta^3(\mathbf{x}),
  \label{eq:inertial-Alice-qubit}
\end{equation}
where $\epsilon_A$ is a dimensionless coupling constant and
\begin{equation}
  c_A(t)=\begin{cases}
    e^{\alpha_A (t-T_A^\mathrm{i})}, & t < T_A^\mathrm{i} \\
    1, & T_A^\mathrm{i} \leq t \leq T_A^\mathrm{f} \\
    e^{-\alpha_A (t-T_A^\mathrm{f})}, & t > T_A^\mathrm{f}
  \end{cases}
  \label{eq:Alice-switching-function}
\end{equation}
models the switching function. Similarly, the function modeling Bob's qubit+field interaction is
\begin{equation}
  f_B(t,\mathbf{x})=\epsilon_Bc_B(t)\delta^3(\mathbf{x}-L\mathbf{\hat{x}}),
  \label{eq:inertial-Bob-qubit}
\end{equation}
where $\epsilon_B$ is a dimensionless coupling constant and $c_B(t)$ is defined as $c_A(t)$ but replacing the $A$'s by $B$'s in Eq.~\eqref{eq:Alice-switching-function}.

Our goal is to explicitly evaluate the classical channel capacity in Eq.~\eqref{eq:classical-channel-capacity} and analyze its dependence on the various parameters involved in this communication process. To this end, we first substitute Eq.~\eqref{eq:inertial-Bob-qubit} in Eq.~\eqref{eq:causal-operator-minkowski} to write
\begin{multline}
  Ef_B(t,\mathbf{x})=\frac{\epsilon_B}{4\pi|\mathbf{x}-L\mathbf{\hat{x}}|}\left[c_B\left(t-|\mathbf{x}-L\mathbf{\hat{x}}|\right)\right. \\
  \left. - c_B\left(t+|\mathbf{x}-L\mathbf{\hat{x}}|\right)\right].
  \label{eq:inertial-EfB}
\end{multline}
Then, by using Eqs.~\eqref{eq:inertial-Alice-qubit} and~(\ref{eq:inertial-EfB}) in Eq.~\eqref{eq:def-nabla}, we can cast the smeared propagator as
\begin{multline}
  \Delta(f_A,f_B) = \frac{\epsilon_A\epsilon_B}{4\pi L}\int_\mathbb{R}dt\,c_A(t)\left[c_B(t-L) -c_B(t+L)\right].
  \label{eq:inertial-Delta}
\end{multline}

Now, note that $\nu_B$ defined in Eq.~\eqref{eq:nu-B-def} depends on Bob's state of motion as well as the quantum field state. Let us consider two cases: if the field is initially in the inertial vacuum state $|0_M\rangle$, $\nu_B$ is simply
\begin{equation}
  \nu_B=\exp\left[-2\langle KEf_B,KEf_B\rangle\right],
  \label{eq:inertial-vacuum-nu-B}
\end{equation}
where $\langle\;,\;\rangle$ is the Klein-Gordon inner product~\eqref{eq:KG-inner-product} and we recall that $K:\mathcal{S}^\mathbb{C}\rightarrow \mathcal{H}$ takes the positive-frequency part of any solution of Eq.~(\ref{eq:minkowski-KG-equation}). Now, if the field is in a KMS (thermal) state at temperature $\Theta$, then~\cite{kay1985a}
\begin{equation}
  \nu_B=\exp\left[-2\left\langle KEf_B,\coth\left(\frac{\beta \hat{h}}{2}\right)KEf_B\right\rangle\right],
  \label{eq:inertial-thermal-nu-B}
\end{equation}
where $\beta\equiv\Theta^{-1}$ is the inverse temperature and ${\hat{h}:\mathcal{H}\rightarrow\mathcal{H}}$ is the  \textit{one-particle Hamiltonian}, which is given by  $\hat{h}=i\partial_t$ and satisfies 
$${H_\phi=d\Gamma(\hat{h})\equiv 1\oplus \hat{h}\oplus \left(\hat{h}\otimes \hat{h}\right)\oplus \cdots}\;.$$ 
We note that, in the zero-temperature limit (i.e., $\beta\rightarrow \infty$), Eq.~(\ref{eq:inertial-thermal-nu-B}) reduces to Eq.~(\ref{eq:inertial-vacuum-nu-B}), as it should be. 

Since the modes $u_\mathbf{k}$ defined in Eq.~\eqref{eq:inertial-modes} form a complete basis for $\mathcal{H}$, we can decompose $Ef_B$ as 
\begin{equation}
    KEf_B=\int d^3{\bf k} \; \langle u_{\bf k}, Ef_B\rangle \; u_{\bf k}
\end{equation}
and thus, as $\hat{h}$ is diagonal in this basis, we can write
\begin{multline}
  \left\langle KEf_B,\coth\left(\frac{\beta \hat{h}}{2}\right)KEf_B\right\rangle  \\
  = \int d^3\mathbf{k} \;\coth\left(\frac{\beta|\mathbf{k}|}{2}\right) \left|\langle u_\mathbf{k},Ef_B\rangle\right|^2.
  \label{eq:thermal-product}
\end{multline}
By making use of Eq.~\eqref{eq:KG-inner-product} and Lemma 3.2.1 of~\cite{wald94} we can cast the Klein-Gordon inner product in Eq.~(\ref{eq:thermal-product}) as 
\begin{equation}
  \begin{aligned}
    \langle u_\mathbf{k},Ef_B\rangle & = i\int_\mathcal{M}\epsilon_{\mathcal{M}}\,\overline{u_\mathbf{k}(x)}f_B(x)
  \end{aligned}
  \label{eq:product-uk-KEfb0}
\end{equation}
which, by using Eq.~(\ref{eq:inertial-modes}), can be put in the form 
\begin{equation}
  \begin{aligned}
    \langle u_\mathbf{k},Ef_B\rangle & = \frac{i\epsilon_B}{2^\frac{3}{2}\pi|\mathbf{k}|^\frac{1}{2}}\,\widetilde{c_B}(|\mathbf{k}|)e^{-ik_xL},
  \end{aligned}
  \label{eq:product-uk-KEfb}
\end{equation}
where  
\begin{equation}
  \widetilde{c_B}(\omega)\equiv\frac{1}{\sqrt{2\pi}}\int_\mathbb{R}dt\,e^{i\omega t}c_B(t)
\end{equation}
is the Fourier transform of $c_B(t)$. Putting together Eqs.~\eqref{eq:inertial-thermal-nu-B},~\eqref{eq:thermal-product}, and~\eqref{eq:product-uk-KEfb} we obtain 
\begin{equation}
  \nu_B (\Theta) =\exp\left[-\frac{2\epsilon_B^2}{\pi}\int_0^\infty dk\,k\,\coth{\left(\frac{k}{2\Theta}\right)}|\tilde{c_B}(k)|^2\right].
  \label{eq:inertial-thermal-nu-B-final}
\end{equation}

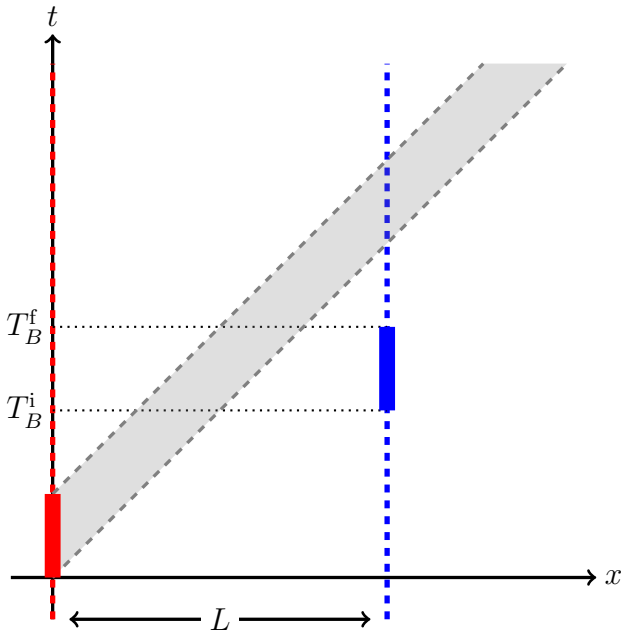
\begin{figure}
  \centering
  \begin{tikzpicture}[scale=0.9\columnwidth/7cm,every label/.append style={text=black,font=\large}]
    %Redimensiona esse tikz picture, que tem 8 cm, para 90% da largura da coluna
    \coordinate (O) at (0,0);
    \coordinate[label=right:$x$] (R) at (6.5,0);
    \coordinate[label=above:$t$] (T) at (0,6.5);
    \coordinate (L) at (-0.5,0);
    \coordinate (B) at (0,-0.5);
    \coordinate (BobB) at (4,-0.5);
    \coordinate (AliceTi) at (0,0);
    \coordinate (AliceTf) at (0,1);
    \coordinate (FP1) at (6.15,6.15);
    \coordinate (FP2) at (5.15,6.15);
    \coordinate[label=left:$T^\mathrm{i}_B$] (TBi) at (0,2);
    \coordinate[label=left:$T^\mathrm{f}_B$] (TBf) at (0,3);
    
    \draw[black,line width=1.25pt,->] (L) -- (R);
    \draw[black,line width=1.25pt,->] (B) -- (T);
    \draw[dashed,red,line width=2pt,domain=-0.5:6.15] plot (0,\x);
    \draw[dashed,blue,line width=2pt,domain=-0.5:6.15] plot (4,\x);
    \draw[black,line width=0.8pt,dotted,domain=0:4] plot (\x,2);
    \draw[black,line width=0.8pt,dotted,domain=0:4] plot (\x,3);
    \draw[dashed,gray,,line width=1.25,domain=0:6.15] plot (\x,\x);
    \draw[dashed,gray,,line width=1.25,domain=0:5.15] plot (\x,\x+1);
    \fill[gray,opacity=0.25]  (AliceTi) -- (AliceTf) -- (FP2) -- (FP1) -- cycle;
    \draw[red,line width=6pt,domain=0:1] plot (0,\x);
    \draw[blue,line width=6pt,domain=2:3] plot (4,\x);
    \path[<->] ($(B)!0.05!(BobB)$) edge[line width=1.25pt] node[fill=white,anchor=center,pos=0.5] {\large $L$} ($(B)!0.95!(BobB)$);
  \end{tikzpicture}
  \caption{Spacetime diagram representing Alice's worldline (red dashed line) and Bob's worldline (blue dashed line). The red and blue rectangles represent the regions where their respective qubits interact with the quantum field. As we are considering a massless field, the gray region represents the region where signals emitted by Alice should be present.}
  \label{fig:inertial-spacetime-diagram}
\end{figure}

We can now use Eqs.~\eqref{eq:classical-channel-capacity},~\eqref{eq:inertial-Delta}, and~\eqref{eq:inertial-thermal-nu-B-final} to  investigate the classical channel capacity when sender and receiver are inertial observers at rest relative to each other. Let us consider that Alice and Bob let their qubits interact with the quantum field for the same amount of time 
$$\Delta T\equiv T_A^\mathrm{f}-T_A^\mathrm{i}=T_B^\mathrm{f}-T_B^\mathrm{i},$$ 
where, for the sake of simplicity, we have set $T_A^\mathrm{i}=0$. We note that by choosing large values of $\alpha_A,\alpha_B$ (i.e., $\alpha_A,\alpha_B \gg 1/\Delta T$) in the switching functions $c_A(t), c_B(t)$, we can model the case where qubit+field interactions take place at finite time intervals $\Delta T$. In  Fig.~\ref{fig:inertial-spacetime-diagram} we plot Alice and Bob worldlines for a spatial separation $L$ as well as the regions where the emission and detection events take place. Note that the emission and detection events are spacelike separated whenever  $T_B^\mathrm{i}<L-\Delta T$ or time-like separated whenever $T_B^\mathrm{i}>L+\Delta T$. As the field is massless, the channel capacity is expected to be zero in such cases since Bob cannot intercept any signal emitted by Alice.

\begin{figure}[tp]
  \includegraphics[width=\columnwidth]{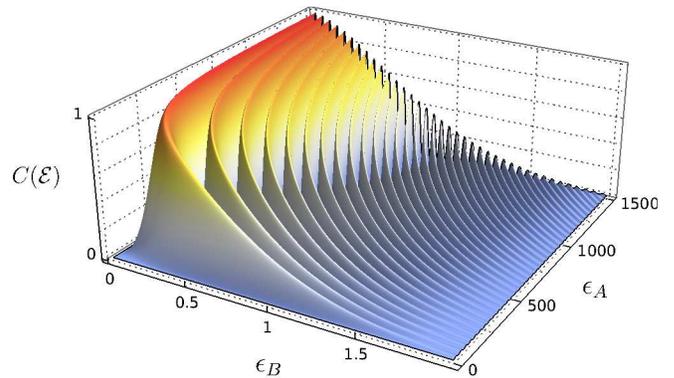}
  \caption{Classical channel capacity as a function of Bob's and Alice's coupling constants, $\epsilon_A$ and $\epsilon_B$, respectively. Here, $\alpha_A=\alpha_B=100\,\Delta T^{-1}$ and $T_B^\mathrm{i}=L=4\,\Delta T$. The channel capacity and coupling constants are dimensionless.}
  \label{fig:inertial-Cxeaxeb}
\end{figure}

Let us begin by analyzing how the coupling constants influence the channel capacity $C\left(\mathcal{E}\right)$. For this purpose we consider the case where the quantum field is initially in the inertial vacuum state and, thus, we are considering the $\beta\rightarrow \infty $ limit in Eq.~(\ref{eq:inertial-thermal-nu-B-final}). In  Fig.~\ref{fig:inertial-Cxeaxeb}, we plot how $C\left(\mathcal{E}\right)$ varies when one changes the couplings $\epsilon_A$ and $\epsilon_B$. For the sake of illustration, we have considered the case where $T_B^\mathrm{i} = L=4\,\Delta T$. This guarantees that Bob's detection process takes place entirely in the gray region of the plot. We can see that the channel capacity increases very close to 1 (maximum efficiency) for large values of $\epsilon_A$ but decreases rapidly with the increase of Bob's coupling constant $\epsilon_B$. This happens, for a fixed value of $\Delta T,$ because as Alice wants to imprint some information on the field state, the stronger her interaction with the field, the more efficient this process will be. On the other hand, Bob's qubit state is altered when it is allowed to interact with the field. If the interaction is too strong (or if it is switched-on for a long period of time), the information encoded in his qubit state can be lost due to quantum decoherence.

\begin{figure}[htp]
  \includegraphics[width=\columnwidth]{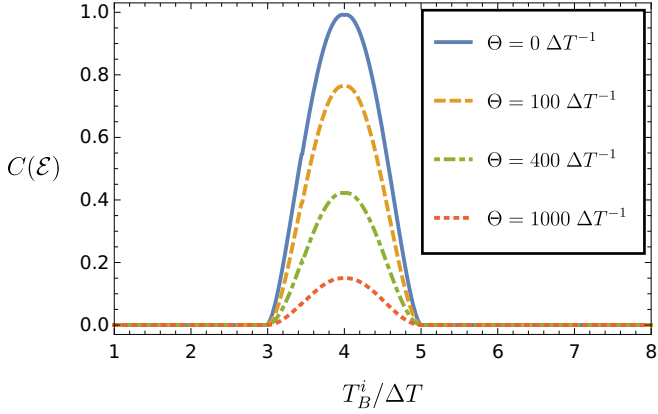}
  \caption{Classical channel capacity as a function of the time $T^\mathrm{i}_B$ when Bob starts the measurement process for different initial thermal states of the field. Each curve represents a different temperature $\Theta$ of the initial quantum field state (with $\Theta=0$ representing the inertial vacuum state). Here, $\alpha_A=\alpha_B=100\,\Delta T^{-1}$, $\epsilon_A=800$, $\epsilon_B=0.05$, and $L=4\,\Delta T$.}
  \label{fig:inertial-CxTbi}
\end{figure}

Having established how one can tune $\epsilon_A$ and $\epsilon_B$ to maximize the channel capacity, let us choose suitable values for the coupling constants and investigate the communication process for different choices of the field initial thermal state as well as  different causal relations between Alice emission and Bob measurement events. The results are shown in Fig.~\ref{fig:inertial-CxTbi}, where we plot the channel capacity, $C\left(\mathcal{E}\right),$ as a function of the time $T^i_B$ where Bob begins his measurement. In view of our previous results, we have chosen $\epsilon_A=800,$ $\epsilon_B=0.05$, and $L= 4\Delta T$.  We can see that the channel capacity vanishes if Bob's qubit interacts with the field too soon ($T_B^\mathrm{i}<L-\Delta T=3\,\Delta T$) or too late ($T_B^\mathrm{i}>L+\Delta T=5\,\Delta T$), regardless of the initial field state. One may observe in Fig.~\ref{fig:inertial-spacetime-diagram} that these are the cases where emission and detection events are spacelike and timelike separated, respectively, and thus Bob cannot intercept any signal emitted by Alice. On the other hand, the maximum communication efficiency is reached when $T_B^\mathrm{i}=L=4\,\Delta T$, since now Bob is able to intercept every signal emitted by Alice. Additionally, note how the temperature of the field state limits the maximum channel capacity. The higher the temperature $\Theta$, the greater the noise in the quantum channel. This increases the quantum decoherence in Bob's qubit state (as the decoherence time decreases) and, thus, it becomes impossible to achieve high efficiency in the communication process.

\subsection{Inertial sender, accelerated receiver}

\begin{figure*}[tp]
  \subfloat[]{\label{figa}
    \begin{tikzpicture}[scale=0.9\columnwidth/6cm,every label/.append style={text=black,font=\large}]
      %Redimensiona esse tikz picture, que tem 6 cm, para 90% da largura da coluna
      \coordinate (O) at (0,0);
      \coordinate[label=right:$x$] (R) at (5.5,0);
      \coordinate[label=above:$t$] (T) at (0,4.5);
      \coordinate (L) at (-0.5,0);
      \coordinate (B) at (0,-1.5);
      \coordinate (AliceTi) at (0,-1);
      \coordinate (AliceTf) at (0,0);
      \coordinate (FP1) at (5.2,4.2);
      \coordinate (FP2) at (4.2,4.2);
      \coordinate[label=below left:{$x_0$}] (x0) at (2,0);
      \draw[black,line width=1.25pt,->] (L) -- (R);
      \draw[black,line width=1.25pt,->] (B) -- (T);
      \draw[dashed,red,line width=2pt,domain=-1.5:4.2] plot (0,\x);
      \draw[dashed,orange,line width=2pt,domain=-1.5:4.2] plot ({2.0+(sqrt(1+(0.1*\x)*(0.1*\x))-1)/0.1},\x);
      \draw[dashed,gray,,line width=1.25,domain=0:5.2] plot (\x,\x-1);
      \draw[dashed,gray,,line width=1.25,domain=0:4.2] plot (\x,\x);
      \fill[gray,opacity=0.25]  (AliceTi) -- (AliceTf) -- (FP2) -- (FP1) -- cycle;
      \draw[red,line width=6pt,domain=-1:0] plot (0,\x);
      \draw[orange,line width=6pt,domain=1.235:2.25] plot ({2.0+(sqrt(1+(0.1*\x)*(0.1*\x))-1)/0.1},\x);
    \end{tikzpicture}
  }\hfill
  \subfloat[]{\label{figb}
    \begin{tikzpicture}[scale=0.9\columnwidth/6cm,every label/.append style={text=black,font=\large}]
      %Redimensiona esse tikz picture, que tem 6 cm, para 90% da largura da coluna
      \coordinate (O) at (0,0);
      \coordinate[label=right:$x$] (R) at (5.5,0);
      \coordinate[label=above:$t$] (T) at (0,4.5);
      \coordinate (L) at (-0.5,0);
      \coordinate (B) at (0,-1.5);
      \coordinate (AliceTi) at (0,-1);
      \coordinate (AliceTf) at (0,0);
      \coordinate (FP1) at (5.2,4.2);
      \coordinate (FP2) at (4.2,4.2);
      \coordinate[label=below left:{$x_0$}] (x0) at (2,0);
      
      \draw[black,line width=1.25pt,->] (L) -- (R);
      \draw[black,line width=1.25pt,->] (B) -- (T);
      \draw[dashed,red,line width=2pt,domain=-1.5:4.2] plot (0,\x);
      \draw[dashed,blue,line width=2pt,domain=-1.5:4.2] plot ({2.0+(sqrt(1+(0.5*\x)*(0.5*\x))-1)/0.5},\x);
      \draw[dashed,gray,,line width=1.25,domain=0:5.2] plot (\x,\x-1);
      \draw[dashed,gray,,line width=1.25,domain=0:4.2] plot (\x,\x);
      \fill[gray,opacity=0.25]  (AliceTi) -- (AliceTf) -- (FP2) -- (FP1) -- cycle;
      \draw[red,line width=6pt,domain=-1:0] plot (0,\x);
      \draw[blue,line width=6pt,domain=1.5:3.204] plot ({2.0+(sqrt(1+(0.5*\x)*(0.5*\x))-1)/0.5},\x);
    \end{tikzpicture}
  }
  \caption{Spacetime diagram, in Cartesian coordinates $(t,x)$, representing Alice (red dashed line) and Bob worldline with different accelerations (blue and orange dashed lines). The solid regions represent emission and detection events that maximize the channel capacity. The detectors remain switched-on for the same proper-time interval.}
\label{fig:accelerated-spacetime-diagrams}
\end{figure*}
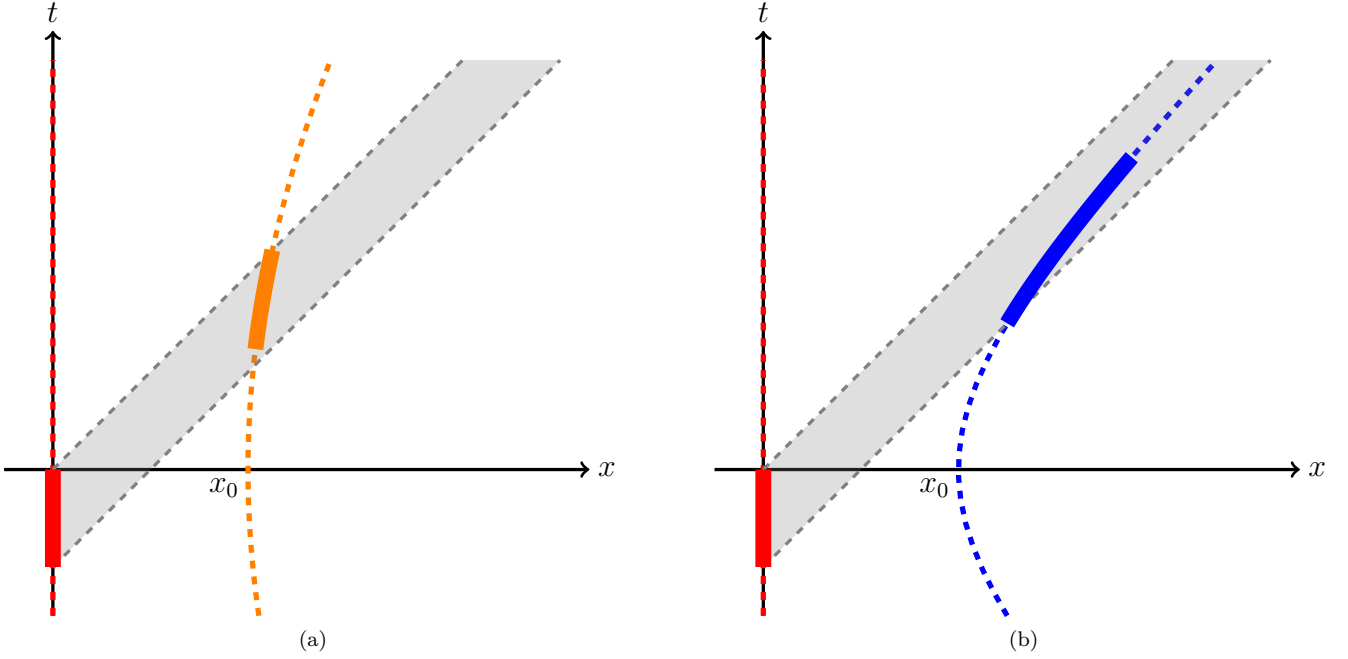

Let us consider now the following scenario: suppose Alice is at rest at the origin of some inertial Cartesian coordinate system $(t,x,y,z)$ and wants to transmit information to Bob, which travels uniformly accelerated following the worldline
\begin{equation}
  \begin{aligned}
    t_B(\tau) & =a^{-1}\sinh\left(a\tau\right), \\
    x_B(\tau) & = x_0 + a^{-1}\left[\cosh\left(a\tau\right)-1\right], \\
    y_B(\tau) & = z_B(\tau) = 0.
  \end{aligned}
\end{equation}
Here, $a$ is Bob proper acceleration, $\tau$ is his proper-time (synchronized as $\tau=0$ when $t=0$), and $x_0$ is the spatial distance between Bob and Alice (as measured by Alice) at the point of maximum approximation. Both worldlines are shown in Fig.~\ref{fig:accelerated-spacetime-diagrams} for two different values of Bob proper acceleration $a$. The quantum field is supposed to be initially in the inertial vacuum state $|0_M\rangle$ and we consider again that both observers are equipped with pointlike detectors which are continuously switched-on/off. The function modeling Alice's qubit+field interaction remains the one in Eq.~\eqref{eq:inertial-Alice-qubit}. 

To discuss Bob's qubit interaction with the field, let us first introduce  Rindler coordinates $(\tau, \xi, y,z)$, with $\tau, \xi \in \mathbb{R}$  implicitly defined by
\begin{equation}
  t=a^{-1}e^{a\xi}\sinh{a\tau}\;,\;\;
  x=\left(x_0-\frac{1}{a}\right)+a^{-1}e^{a\xi}\cosh{a\tau}.\\
  \label{rindlercoord}
\end{equation}
These coordinates cover the right Rindler wedge (RRW), i.e., the region defined by $\left[x-(x_0-1/a)\right]>|t|$, in which the metric takes the form
\begin{equation}
  g=e^{2a\xi}(-d\tau\otimes d\tau +d\xi\otimes d\xi )+dy\otimes dy +dz\otimes dz.
\end{equation}
In such coordinates, Bob remains static at $\xi=y=z=0$ and its qubit+field interaction is simply described by the function
\begin{equation}
  f_B(\tau,\xi,y,z)=\epsilon_B\,c_B(\tau)\,\delta(\xi)\,\delta(y)\,\delta(z),
  \label{eq:accelerated-Bob-qubit}
\end{equation}
where
\begin{equation}
  c_B(\tau)=\begin{cases}
    e^{\alpha_B (\tau-\tau_B^\mathrm{i})}, & \tau < \tau_B^\mathrm{i} \\
    1, & \tau_B^\mathrm{i} \leq \tau \leq \tau_B^\mathrm{f} \;.\\
    e^{-\alpha_B (\tau-\tau_B^\mathrm{f})}, & \tau > \tau_B^\mathrm{f}
  \end{cases}
\end{equation}

In order to analyze the channel capacity, we need to evaluate again $\Delta(f_A,f_B)$. By an analogous procedure to the one leading to Eq.~\eqref{eq:inertial-EfB}, we obtain
\begin{equation}
  Ef_A(t,\mathbf{x})=\frac{\epsilon_A}{4\pi|\mathbf{x}|}\left[c_A\left(t-|\mathbf{x}|\right) - c_A\left(t+|\mathbf{x}|\right)\right].
  \label{eq:inertial-EfA}
\end{equation}
Using Eq.~\eqref{eq:covariant-CCR}, we have
\begin{equation}
  \begin{aligned}
    \Delta(f_A,f_B) &= -\Delta(f_B,f_A) \\
                    &=- \int_\mathcal{M} \epsilon_{\mathcal{M}}\;f_B(x)Ef_A(x)
  \end{aligned}
\end{equation}
which, in Rindler coordinates~(\ref{rindlercoord}), can be  straightforwardly evaluated giving
\begin{multline}
  \Delta(f_A,f_B)= \frac{\epsilon_A\epsilon_B}{4\pi} \int_\mathbb{R}d\tau \; \frac{c_B(\tau)}{|x_B(\tau)|}\left\{c_A\left[t_B(\tau)+x_B(\tau)\right]\right. \\
  \left.-c_A\left[t_B(\tau)-x_B(\tau)\right]\right\}.
  \label{eq:accelerated-Delta-fA-fB}
\end{multline}

Now, we proceed to calculate the quantity $\nu_B$ defined in Eq.~\eqref{eq:nu-B-def}, which depends on Bob's state of motion and the initial state of the quantum field. In order to do so, it will be useful to introduce the so-called Right Rindler modes defined by
\begin{equation}
  v_{\omega \mathbf{k_\perp}}^R\equiv\left[\frac{\sinh(\pi\omega/a)}{4\pi^4a}\right]^{1/2}K_{i\omega/a}\left(\frac{|\mathbf{k_\perp}|e^{a\xi}}{a}\right)e^{i\mathbf{k_\perp\cdot x_\perp}}e^{-i\omega\tau}
  \label{vR}
\end{equation}
in the RRW and vanishing in the left Rindler wedge (LRW), which is the region where $\left[x-(x_0-1/a)\right]>-|t|$. Here, $\mathbf{x_\perp}\equiv(y,z)$, $\mathbf{k_\perp}\in\mathbb{R}^2$,  $\omega>0$, and $K_{\nu}(x)$ is the modified Bessel function. The left Rindler modes, $v^L_{\omega {\bf k}_\perp}$, are defined by $v^L_{\omega {\bf k}_\perp}(t,x,{\bf x}_\perp)\equiv \overline{v^R_{\omega {\bf k}_\perp}}(-t,-x,{\bf x}_\perp)$. Hence, they vanish in the RRW and take the form~(\ref{vR}) in Rindler coordinates covering the LRW.

By using $v^R_{\omega {\bf k}_\perp}$ and $v^L_{\omega {\bf k}_\perp}$ we can define the so-called Unruh modes 
\begin{equation}
  w^1_{\omega \mathbf{k_\perp}}\equiv\frac{v_{\omega \mathbf{k_\perp}}^{R}+e^{-\pi\omega/a}\,\overline{v_{\omega -\mathbf{k_\perp}}^{L}}}{\sqrt{1-e^{-2\pi\omega/a}}},\label{w1}
\end{equation}
\begin{equation}
  w^2_{\omega \mathbf{k_\perp}}\equiv\frac{v_{\omega \mathbf{k_\perp}}^{L}+e^{-\pi\omega/a}\,\overline{v_{\omega -\mathbf{k_\perp}}^{R}}}{\sqrt{1-e^{-2\pi\omega/a}}},
\end{equation}
which have purely positive frequency relative to inertial time and comprise a complete basis for the one-particle space $\mathcal{H}$ defined below Eq.~\eqref{SigmaInert}. Therefore, we can write
\begin{eqnarray}
  \langle KEf_B,KEf_B\rangle =\int_{\mathbb{R}} d\omega\int_{\mathbb{R}^2} d^2\mathbf{k_\perp}\;\left|\langle w^1_{\omega \mathbf{k_\perp}},Ef_B\rangle\right|^2,  \nonumber \\
  \label{eq:accelerated-KEfb-norm}
\end{eqnarray}
where we have used that $ w^1_{-\omega \mathbf{k_\perp}}= w^2_{\omega \mathbf{k_\perp}}$. Analogously to Eq.~\eqref{eq:product-uk-KEfb}, we can write  
\begin{equation}
  \langle w^1_{\omega \mathbf{k_\perp}},Ef_B\rangle = i\int \epsilon_{\mathcal{M}}\; \overline{w^1_{\omega \mathbf{k_\perp}}}(x)f_B(x) 
\end{equation}  
which, by using Eq.~(\ref{w1}), gives 
\begin{eqnarray}
 \langle w^1_{\omega \mathbf{k_\perp}},Ef_B\rangle &=& \frac{i\epsilon_B}{\sqrt{1-e^{-2\pi\omega/a}}}\left[\frac{\sinh(\pi\omega/a)}{2\pi^3a}\right]^{1/2} \nonumber \\
  &\times& e^{-\pi\omega/a}K_{i\omega/a}\left(|\mathbf{k_\perp}|/a\right)\widetilde{c_B}(\omega),
  \label{eq:w-KEfb-product}
\end{eqnarray}    
 Substituting Eqs.~\eqref{eq:w-KEfb-product} and~\eqref{eq:accelerated-KEfb-norm} in Eq.~\eqref{eq:inertial-vacuum-nu-B} and evaluating the transverse,  $\mathbf{k_\perp}$, integral by means of the identity
 \begin{equation}
     \int_0^\infty dx x |K_{i\omega/a}(x)|^2= \frac{\omega}{4a \sinh{(\pi \omega/a)}} 
 \end{equation}
 we obtain
\begin{equation}
  \nu_B =\exp\left[-\frac{2\epsilon_B^2}{\pi}\int_0^\infty d\omega\,\omega\coth{\left(\frac{\pi\omega}{a}\right)}|\widetilde{c_B}(\omega)|^2\right].
  \label{eq:accelerated-nu-B-final}
\end{equation}

Now, we can use Eqs.~\eqref{eq:classical-channel-capacity},~\eqref{eq:accelerated-Delta-fA-fB},~and~\eqref{eq:accelerated-nu-B-final} to  investigate the classical channel capacity $C\left(\mathcal{E}\right)$ for inertial sender and accelerated receiver. We consider that Alice and Bob let their qubits interact with the quantum field for the same amount of their respective proper-time, hence 
\begin{equation}
\Delta T \equiv T_A^\mathrm{f}-T_A^\mathrm{i}=\tau_B^\mathrm{f}-\tau_B^\mathrm{i}.
\end{equation} 
In Fig.~\ref{fig:accelerated-spacetime-diagrams}, we plot their spacetime trajectories as well as regions where emission and detection events may take place.
\begin{figure}[tp]
  \includegraphics[width=\columnwidth]{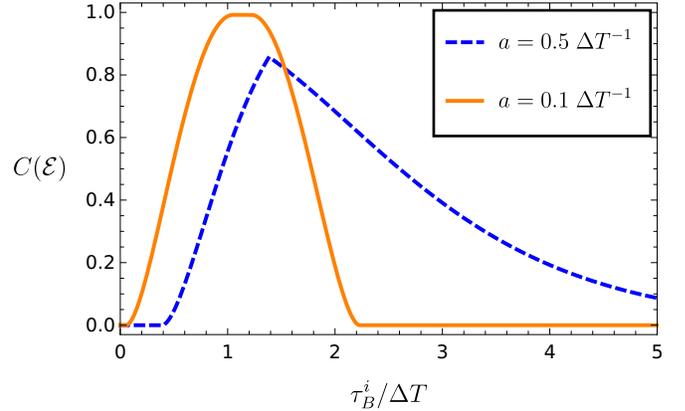}
  \caption{Classical channel capacity as a function of the proper time $\tau_B^\mathrm{i}$ when Bob starts the measuring process, for different proper accelerations. Each curve represents one of the situations schematized in Fig.~\ref{fig:accelerated-spacetime-diagrams}. Here, $\alpha_A=\alpha_B=100\,\Delta T^{-1}$, $\epsilon_A=420$, $\epsilon_B=0.05$ and $x_0=2\,\Delta T$. The field is initially in the inertial vacuum state.}
  \label{fig:accelerated-CxTbi}
\end{figure}
In Fig.~\ref{fig:accelerated-CxTbi} we show the behavior of the classical channel capacity as a function of the (proper) time $\tau^i_B$ in which bob begins the measurement process. Let us first consider the case where that Bob's worldline is the one in Fig.~\ref{figa}, where $x_0=2\,\Delta T$ and $a=0.1\,\Delta T^{-1}$. In this case, Bob intercepts the first and final signal emitted by Alice at proper times $\tau_1\simeq1.05\,\Delta T$ and $\tau_2\simeq2.23\,\Delta T$, respectively. As it can be seen in Fig.~\ref{fig:accelerated-CxTbi}, information exchange is possible only if Bob starts the measurement process at proper-time $\tau_B^\mathrm{i}$ satisfying $\tau_1-\Delta T < \tau_B^\mathrm{i} < \tau_2$. When $\tau_B^\mathrm{i}<\tau_1-\Delta T$ or  $\tau_B^\mathrm{i}>\tau_2$, the classical channel capacity vanishes since the emission and detection events will be spacelike or timelike separated, respectively.
 
Let us consider now that Bob's trajectory is the one depicted in  Fig.~\ref{figb}, where $x_0=2\,\Delta T$ and $a=0.5\,\Delta T^{-1}$. In this case, Bob's worldline intercepts the first signal emitted by Alice at proper time $\tau_1'\simeq1.37\,\Delta T$. In Fig.~\ref{fig:accelerated-CxTbi}, we see that it is exactly when $\tau_B^\mathrm{i}=\tau_1'$ that the maximum channel capacity is attained. However, since the surface $t = x$ is Bob's causal horizon, he never leaves the lightcone section where the information emitted by Alice is traveling. Thus, the channel capacity never vanishes. It only decreases as Bob accelerates away from Alice.

%%%%%%%%%%%%%%%%%%%%%%%%%%%%%%%%%%%%%%%%%%%%%%%%%%%%%%%
\subsection{Work for switching on/off the detectors}
%%%%%%%%%%%%%%%%%%%%%%%%%%%%%%%%%%%%%%%%%%%%%%%%%%%%%%%
We have shown in Sec.~\ref{sec:energy} that one can use the quantum channel presented here to convey arbitrary amounts of information without any extra energy cost. However, some work is necessary to switch-on/off the interaction of each qubit with the quantum field,  which is given by Eq.~\eqref{eq:W-j}. Let us now estimate this energy cost for by means of the inertial detectors previously discussed.

For this purpose, let us first  relate the ordinary Minkowski QFT construction presented at the beginning of this section to the null-surface construction introduced in Sec.~\ref{sec:nullquant}. By recalling that $(t,x,y,z)$ are inertial Cartesian coordinates where Minkowski metric takes the form given in Eq.~(\ref{eta}), let
\begin{equation}
    u\equiv t-|\mathbf{x}|\; {\rm and} \;  v\equiv t+|\mathbf{x}|
    \label{uvcoord}
\end{equation} 
be the retarded and advanced null coordinates, respectively, where $\mathbf{x}\equiv (x,y,z)$.  By defining 
\begin{equation}
    \tan{V}\equiv v,
\end{equation} 
one can cast Eq.~(\ref{eta}) using $u$ and $V$ coordinates as
\begin{equation}
  \eta=-\frac{1}{2}\sec^2V\left(du\otimes dV+ dV\otimes du\right)+\frac{(\tan V - u)^2}{4}g_{\mathbb{S}^2},
\end{equation}
where $g_{\mathbb{S}^2}$ is the standard metric on a unit 2-sphere. 

Now, it is easy to see that Minkowski spacetime is future null asymptotically flat in the following sense~\cite{HE}: there exists a second spacetime $(\widehat{\mathcal{M}},\hat{g})$ with metric
\begin{equation}
  \hat{g}=-\frac{2}{(\sin V-u\cos V)^2} \left(du\otimes dV+ dV\otimes du\right) + g_{\mathbb{S}^2}
\end{equation}
such that: \textbf{(1)} $\mathcal{M}$ is conformally embedded in $\widehat{\mathcal{M}}$ satisfying $\hat{g}=\Omega^2 \eta$ with a conformal factor
\begin{equation}
  \Omega\equiv \frac{2}{\tan V - u};
  \label{eq:conformal-factor-minkowski}
\end{equation}
\textbf{(2)} the future null infinity $\mathcal{I}^+$ is the 3-dimensional null hypersurface  defined by the set
$$\left\{p\in \widehat{\mathcal{M}}|V=\frac{\pi}{2}\right\}$$ 
in which the conformal factor satisfies $\left.\Omega\right|_{\mathcal{I}^+}=0$ and $\left.d\Omega\right|_{\mathcal{I}^+}\neq 0$.

The conformal metric $\hat{g}$ restricted to $\mathcal{I}^+$ takes the form
\begin{equation}
  \hat{g}|_{\mathcal{I}^+}=-\frac{1}{2}\left(du\otimes d\Omega + d\Omega \otimes du \right) +g_{\mathbb{S}^2},
\end{equation}
which has the form given in  Eq.~\eqref{eq:metric-restricted-to-horizon}. Thus, we can perform the null quantization procedure at future null infinity $\mathcal{I}^+$ described in Sec.~\ref{sec:nullquant} with $\lambda=u$.

Now, let us consider the inertial qubit carried by Alice whose interaction with the field is described by the function $f_A=\epsilon_A c_A(t)\delta^3({\bf x})$ with $c_A$ given in Eq.~\eqref{eq:inertial-Alice-qubit}. To compute the energy cost $W_A$, given in Eq.~\eqref{eq:W-j}, to switch-on/off the qubit, we need to first compute  $Ef_A^{\mathfrak{h}}$. To this end, we can use  the explicit form of $Ef_A$ given in Eq.~\eqref{eq:inertial-EfA} together with Eq.~(\ref{uvcoord}) to write
\begin{equation}
  Ef_A = \frac{\epsilon_A}{4\pi|\mathbf{x}|}\left[c_A(u)-c_A(v)\right].
\end{equation}
 By using the above equation, the extension $Ef_A^{\mathfrak{h}}$ of $Ef_A$ to $\mathcal{I}^+$ is simply computed using
\begin{equation}
    Ef_A^{\mathfrak{h}}  \equiv \lim_{v\rightarrow\infty}\; \Omega^{-1} Ef_A 
\end{equation}
and 
$$|{\bf x}|=\frac{v-u}{2}=\Omega^{-1}$$
yielding 
\begin{equation}
  Ef_A^{\mathfrak{h}} = \frac{\epsilon_A}{4\pi}\,c_A(u).
  \label{Efh}
\end{equation}
By using Eqs.~(\ref{Efh}) and~(\ref{eq:inertial-Alice-qubit}) in Eq.~\eqref{eq:W-j} leads to
\begin{equation}
  \begin{aligned}
    W_A & =\int_{\mathcal{I}^+} du\wedge\epsilon_{\mathbb{S}^2}\,\left(\frac{\epsilon_A}{4\pi}\right)^2\left[\partial_u c_A(u)\right]^2 \\
        & =\frac{\epsilon_A^2\alpha_A}{4\pi}.
  \end{aligned}
\end{equation}

We can see that the work necessary to switch-on/off the detector increases with the coupling strength and it is inversely proportional to the time-scale $\tau_{\alpha_A}\equiv \alpha_A^{-1}$ characterizing the switching process. This gives the usual trade-off between the energy  $W_A$ of the process and its characteristic time $\tau_{\alpha_A}$:
\begin{equation}
W_A \tau_{\alpha_A} =\frac{\epsilon_A^2}{4\pi},    
\end{equation}
which shows that the more energy is needed the more rapid is the switching-on/off of the qubit interaction. 

%%%%%%%%%%%%%%%%%%%%%%%%%%%%%%%%%%%%%%%%%%%%%%%%%%%%%%%%%%%%%%%%%%
\section{Conclusions}
\label{sec:finalremarks}

In the present paper, we have analyzed the energy cost in conveying classical and quantum information between two arbitrary observers in asymptotically flat and globally hyperbolic spacetimes (possibly containing black holes) when they use a quantum scalar field as a communication channel. We have shown that the energy variation of the total 2-qubits+field system, $\Delta E$, can be cast as $\Delta E =W_\phi + W_A + W_B + W_{AB}$. This shows that such energy variations has three may contributions: {\bf (1)} $W_\phi$ which accounts for the particle creation due to the change of the spacetime geometry; {\bf (2)} $W_A+ W_B$ which gives the energy needed to switch-on/off qubits $A$ and $B$ used by Alice and Bob, respectively, to communicate; {\bf (3)} $W_{AB}$ which describes the extra energy cost needed for the communication process. We have shown that, by suitably choosing Bob's initial (ready-to-measure) state, the term $W_{AB}$ vanishes. Such condition is satisfied by the channel $\mathcal{E}$ considered here. 

We have then illustrated the communication process and analyzed how the classical channel capacity $C\left(\mathcal{E}\right)$ (and, as a result,  the entanglement-assisted classical and quantum capacities as well) behaves in two paradigmatic examples in Minkowski spacetime: {\bf (A)} when  sender and receiver are inertial observers and {\bf (B)} when the sender is inertial and the receiver is uniformly accelerated. By using example {\bf (A)}, we were able to analyze how the coupling constants as well as the initial field state influence  $C\left(\mathcal{E}\right)$. Example {\bf (B)} enabled us to analyze how causal horizons affects the communication process when the field state is the inertial vacuum (which, by the Unruh effect, is perceived as a thermal state with temperature $T_U=a/2\pi$ by the uniformly accelerated receiver). 

We have ended the paper using the behaviour of Alice inertial qubit in Minkowski spacetime to estimate the energy cost in switching-on/off the interaction, i.e., $W_A$. We have shown that, as one would expect, $W_A$ satisfies the energy-time relation: $W_A \tau_{\alpha_A}=\epsilon_A^2/ 4\pi$, where $\tau_{\alpha_A}$ is the characteristic time of the switching-on/off process and  $\epsilon_A$ is the coupling constant describing Alice's qubit interaction with the field $\phi$. Hence, one would expect to spend an amount $W_X \sim \epsilon_X^2 \tau_{\alpha_X}^{-1}$, $X=A,B$, of energy in order to create the qubits and switch them on/off to perform some task. However, if one has already created the qubits for some purpose (and the energy cost for it is accounted by $W_A+W_B$), there is no extra energy cost in using them to convey information. 
%%%%%%%%%%%%%%%%%%%%%%%%%%%%%%%%%%%%%%%%%%%%%%%%%%%%%%%%%%%%%%%%%%

\acknowledgments

I. B. and A. L. were fully and partially supported by S\~ao Paulo Research Foundation under Grants 2018/23355-2 and 2017/15084-6, respectively.

\end{document}